\documentclass[journal]{IEEEtran}

\usepackage{booktabs}   
\usepackage{cite}

\ifCLASSINFOpdf

\else

\fi

\usepackage[cmex10]{amsmath}
\usepackage{graphicx}
\graphicspath{{eps/}}
\DeclareGraphicsExtensions{.eps}

\usepackage[dvips]{epsfig}
\usepackage[T1]{fontenc}

\usepackage[cmex10]{amsmath}
\usepackage{amssymb}
\usepackage{cite}
\usepackage{lettrine}
\usepackage{color}
\usepackage{amsmath} 
\usepackage{amssymb}  

\begin{document}

\title{High-Throughput Cooperative Communication with Interference Cancellation for Two-Path Relay in Multi-source System}


\author{Hao~Lu,
        Peilin~Hong,
        and~Kaiping~Xue, {\em Member,~IEEE}
        \thanks{
        Manuscript received July 25, 2012; revised October 23, 2012 and March 17, 2013.

        Hao~Lu, Peilin~Hong, and~Kaiping~Xue are with the information Network Lab of EEIS Department, University of Science and Technology of China, Hefei 230027,China (Email: lluhao@mail.ustc.edu.cn, plhong@ustc.edu.cn, kpxue@ustc.edu.cn ).}
        }

\maketitle
\begin{abstract}
Relay-based cooperative communication has become a research focus in recent years because it can achieve diversity gain in wireless networks. In existing works, network coding and two-path relay are adopted to deal with the increase of network size and the half-duplex nature of relay, respectively. To further improve bandwidth efficiency, we propose a novel cooperative transmission scheme which combines network coding and two-path relay together in a multi-source system. Due to the utilization of two-path relay, our proposed scheme achieves full-rate transmission. Adopting complex field network coding (CFNC) at both sources and relays ensures that symbols from different sources are allowed to be broadcast in the same time slot. We also adopt physical-layer network coding (PNC) at relay nodes to deal with the inter-relay interference caused by the two-path relay. With careful process design, the ideal throughput of our scheme achieves by 1 symbol per source per time slot (sym/S/TS). Furthermore, the theoretical analysis provides a method to estimate the symbol error probability (SEP) and throughput in additive complex white Gaussian noise (AWGN) and Rayleigh fading channels. The simulation results verify the improvement achieved by the proposed scheme.
\end{abstract}



%
\IEEEpeerreviewmaketitle

\section{Introduction}

\IEEEPARstart{B}{}ased on the fact that it is difficult to pack multiple antennas per terminal due to the limit of physical space, relay-based cooperative communication is especially eye-catching since it can extend coverage and achieve spatial diversity gain in wireless communication networks. However, due to the increase of network size and the half-duplex nature of relay, traditional relay schemes are bandwidth inefficient. Two methods were proposed to break through this bandwidth bottleneck:

1) Network coding. Physical-layer network coding (PNC) was proposed in [1] to map superposition of electromagnetic signals to simple Galois field GF($2^{n}$) additions of digital bit streams. In [2], Galois field network coding (GFNC) was applied to wireless communication networks to enhance bandwidth efficiency.
\begin{figure}[h]
\centering

\includegraphics[width=0.45\textwidth]{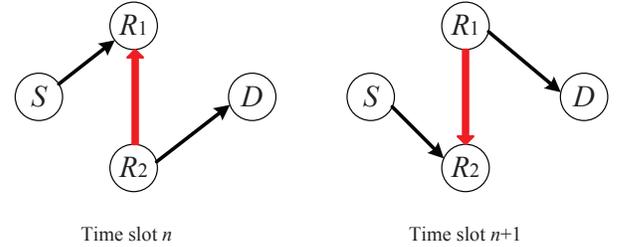}
\caption{Two-path relay scheme model}

\label{f1}
\end{figure}
Afterwards, Wang and Giannakis put forward a novel cooperative approach based on complex field network coding (CFNC) [3], which makes it possible for sources to broadcast symbols in the same time slot (TS) on the same resource blocks (RBs) after symbol-level operations at the physical layer. Deploying CFNC, a cooperative network with $N_{S}$ sources offers throughput as high as $1/2$ symbol per source per time slot (sym/S/TS), relative to $1/(2N_{S})$ sym/S/TS in traditional cooperative system and $1/(N_{S}+1)$ sym/S/TS in GFNC-based cooperative system. Hence, CFNC has been applied to diverse cooperative networks [4]-[7].

2) Two-path relay. Relays usually work in the half-duplex mode, where a transmission process often requests two TSs for the relay to receive the signal and then forward it respectively, which results in a loss in spectral efficiency. To overcome it, in [8] and [9], a spectral efficient cooperative scheme named two-path successive relay was proposed. As illustrated in Fig. 1, at each TS, the source $S$ sends its symbol to one of the relays and the other relay forwards the symbol received from $S$ at the previous TS. Therefore, only $(L+1)$ TSs are required to transmit $L$ symbols from $S$ to $D$. However, two-path relay scheme results in inter-relay interference (IRI), which is marked as the red lines in Fig.$\:$1. As long as the IRI can be canceled, we can double the transmission rate without much sacrifice of the performance in terms of symbol error probability (SEP). In [10], network coding at relay nodes was applied to achieve inter-relay interference cancellation (IRIC). In [11], full interference cancellation (FIC) was proposed, which requires complicated detection. The orthogonality of the real part and the imaginary part of the symbol was utilized to distinguish the information and interference in [12]. More research on two-path relay in recent years can be found in [13][14].


Both of the methods above have improved bandwidth efficiency and throughput of relay-based communication. Network coding deals with the increase of network size while two-path relay overcomes the half-duplex nature of relay nodes. Although CFNC allows simultaneous transmissions from all sources, $2L$ TSs are required to complete the transmission of $L$ symbols for each source due to the half-duplex nature of relay. On the other hand, existing researches on two-path relay only consider single-source models. With the increase of network size, the single-source models in [10]-[14] cannot be directly generalized to multi-source systems with simultaneous transmissions at sources.

Therefore, in order to further improve bandwidth efficiency, we propose a novel transmission scheme based on two-path relay in multi-source wireless networks. Our scheme combines network coding and two-path relay together in a (multi-source,2-relay,1-destination) system. To allow simultaneous transmissions of all sources, CFNC is operated before transmission. With the help of two-path relay, full-rate transmission of multi-source system is achieved. We use physical-layer network coding at relay nodes to deal with the IRI caused by two-path relay and the interference is successfully canceled at destination. Only $L+1$ TSs are needed to send $L$ symbols for each source. Compared with CFNC, our proposed cooperative transmission in $(N_{S},2,1)$ system achieves ideal throughput as high as 1 sym/S/TS, twice as high as that of CFNC in [3]. Compared with [10]-[14], multi-source model, rather than single-source, is introduced to two-path relay system without requirements for additional TSs and resource blocks. Besides, theoretical analysis and simulation results also demonstrate the improvement of our scheme's performance.

The rest of this paper is organized as follows: Section II depicts the system model. Section III details the proposed transmission scheme and Section IV illustrates the IRIC algorithm. The theoretical analysis of the SEP and the throughput are presented in Section V. In Section VI, simulation results are given to verify the improvement of throughput in different scenarios. Finally, Section VII summarizes the paper.



\section{System Model}
As illustrated in Fig.~2, we consider a two-path relay cooperative system that consists of multi-sources ($S_{1},S_{2},...,S_{N_{S}}$), two relays ($R_{1},R_{2}$) and one destination ($D$). One of the applications is being adopted by the uplinks in cellular networks to promote transmission reliability for cell-edge users, shown in Fig.$\:$2(b). Apart from extending coverage and achieving diversity gain, relay-based cooperative scheme could also reduce the required transmitted power at sources, which leads to lower inter-cell interference. Due to lower transmitted power at sources, we consider that the direct links between sources and destination ($S\text{\bfseries{--}}D$) do not exist. All sources broadcast their own modulated symbols on the same RB in the same TS. Two relay nodes, working in the half-duplex mode, take turns to detect-and-forward (DF) received symbols to destination. Relay nodes also employ the same RB, which results in the IRI.
\begin{figure}[h]
\centering
\includegraphics[width=0.48\textwidth]{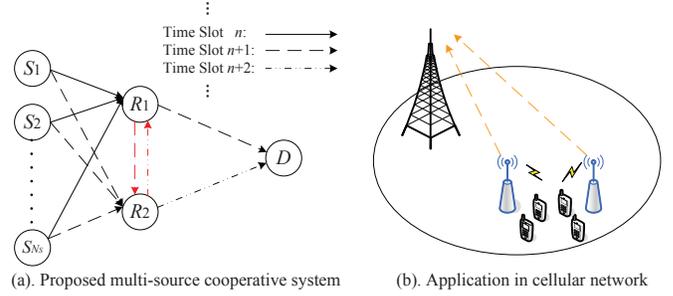}
\caption{System model}
\end{figure}
\begin{table}[ht]

\centering
\caption{Notations}
 \begin{tabular}{lp{6cm}}
 \toprule
  Notations & Description\\
  \midrule
  $N_{S}$ & the number of sources \\
  $S_{i}$ & the $i$th source, $i=1,2,...,N_{S}$\\
  $R_{m}$ & the $m$th relay, $m=1,2$\\
  $D$ & the destination node\\
  $h_{link}$ & channel coefficient, including: $h_{S_{i}R_{m}}$ ($S\text{\bfseries{--}}R$), $h_{12}$ ($R\text{\bfseries{--}}R$), $h_{R_{m}D}$ ($R\text{\bfseries{--}}D$)\\
  $w$ & AWGN, including: $w_{R_{m}}$ (received by $R$), $w_{D}$ (received by $D$)\\

  $(\cdot)^{T}$ & transpose \\

  $\Theta^{T}$ & $[\theta_{1},\theta_{2},...,\theta_{N_{S}}]$, precoding vector adopted by CFNC  \\
  $\mathbf{x}$ & $[x_{1},x_{2},...,x_{N_{S}}]^{T}$, sources' original symbols  \\
  ${\mathbf{x}}_{r}$ & $[x_{1_{r}},x_{2_{r}},...,x_{N_{S_{r}}}]^{T}$, symbols sent by relay before multiplied by $\Theta^{T}$ \\
  $x_{i}^{(n)}$ & the symbol sent by source $i$ at time slot $n$ \\
  $x_{i_{r}}^{(n)}$ & the symbol sent by relay at time slot $n$ corresponding to source $i$\\
  $\hat{x}$ & the estimation of $x$ \\
  $y_{R_{m}}^{(n)},y_{D}^{(n)}$ & the symbol received by $R_{m}$ and $D$ respectively at TS $n$ \\
  $f$, $g$ & the PNC mapping function \\
  $\Re\{x\}$, $\Im\{x\}$ & the real part and imaginary part of $x$ \\

  $M$ & the cardinality of modulation symbol set \\
  $j$ & unit imaginary number \\
  $s$ & $\Theta^{T}\mathbf{x}$, the symbol from sources after CFNC\\
  $\mathbf{z}$ & $\mathbf{x}+{\mathbf{x}}_{r}$, the superposition of desired symbol $\mathbf{x}$ and interference ${\mathbf{x}}_{r}$ \\
  $y$ & $\Theta^{T}\mathbf{z}$, the symbol received by $R$ in AWGN channels regardless of noise \\
  $\tilde{\mathbf{z}}$ & ${\mathbf{H}}_{SR}{\mathbf{x}}+h_{12}{\mathbf{x}}_{r}$, the superposition of desired symbol ${\mathbf{H}}_{SR}{\mathbf{x}}$ and interference $h_{12}{\mathbf{x}}_{r}$ \\
  $\tilde{y}$ & $\Theta^{T}\tilde{\mathbf{z}}$, the symbol received by $R$ in flat fading channels regardless of noise \\
  ${\mathcal{A}}_{x},{\mathcal{A}}_{s},{\mathcal{A}}_{z},\cdots$ & the modulation symbol set of $x,s,z,\cdots$ \\
  $\mathcal{CN}$$(0,\sigma^{2})$  & the circular symmetric complex Gaussian distribution with zero mean and variance $\sigma^{2}$ \\
  $Q(x)$ & $1/\sqrt{2\pi}\int_{x}^{\infty}\textrm{exp}(-t^{2}/2)\mathrm{d}t$, the Gaussian tail function  \\
  $P_{e}$ & probability of error \\
  $\mathrm{E}\{\cdot\}$ & the mathematical expectation \\
  $N_{0}$ & the noise power spectral density \\
  $T$ & throughput \\
  $\angle h$ & the phase of $h$ \\
  \bottomrule
 \end{tabular}

\end{table}
Channel coefficients between $S_{i}$ and $R_{m}$, $R_{1}$ and $R_{2}$, $R_{m}$ and $D$ are denoted by $h_{S_{i}R_{m}}$, $h_{12}$ and $h_{R_{m}D}$ where $i=1,2,...,N_{S}$ and $m=1,2$. According to [15], when the coherence time of the channel is much greater than the symbol period, channel reciprocity (CR) can theoretically be assumed for channels in which uplink and downlink transmissions share the same frequency spectrum. Therefore, regardless of relay nodes' mobility, we assume ${\mathrm{E[}}|h_{12}|^{2}{\mathrm{]}}={\mathrm{E[}}|h_{21}|^{2}{\mathrm{]}}=\sigma^{2}_{12}$. The corresponding channel noise $w_{R_{m}}$ and $w_{D}$ are assumed to be additive complex white Gaussian noise (AWGN) with mean zero and variance $N_{0}$. All channel coefficients and noise are supposed to remain constant within one TS and vary independently from TS to TS. Throughout this paper, we mainly consider two kinds of channels: ideal AWGN channels with all channel coefficients being 1 and flat slow fading channels. Under normal circumstances, each receiver knows the channel state information (CSI) ($R_{m}$ knows $h_{S_{i}R_{m}}$ and $h_{12}$, $D$ knows $h_{R_{m}D}$).

In addition, we introduce some primary notations to describe our system, shown in Table I.



\section{The Proposed Transmission Scheme}
Under the assumption that $R_{1}$ and $R_{2}$ are assigned to relay messages for $N_{S}$ sources, each source has continuous $L$ symbols required to send to destination.
To guarantee that different sources' symbols can be separated, CFNC is adopted here to establish a one-to-one correspondence between $\mathbf{x}$$=[x_{1},x_{2},...,x_{N_{S}}]^{T}$ and $\Theta^{T}\mathbf{x}$ where $\Theta^{T}=[\theta_{1},\theta_{2},...,\theta_{N_{S}}]$. Before transmission, the source symbol $x_{i}$ from $S_{i}$ is multiplied by $\theta_{i}$, $i=1,2,...,N_{S}$. The design of $\Theta^{T}$ is supposed to satisfy $\Theta^{T}{\mathbf{x}}\neq \Theta^{T}{\mathbf{x}}'$ if ${\mathbf{x}}\neq{\mathbf{x}}'$, which allows simultaneous transmissions from multiple sources. Among the different choices for $\Theta^{T}$, we take it to be any row of the Vandermonde matrix:
\begin{displaymath}
\left[ \begin{array}{cccc}
1 & \delta_{1} & \ldots & \delta^{N_{S}-1}_{1} \\
1 & \delta_{2} & \ldots & \delta^{N_{S}-1}_{2}\\
\vdots & \vdots & &\vdots\\
1 & \delta_{N_{S}} & \ldots & \delta^{N_{S}-1}_{N_{S}}\\
\end{array} \right]_{N_{S}\times N_{S}}
\end{displaymath}
where $\{\delta_{u}\}^{N_{S}}_{u=1}=e^{j\pi (4u-1)/(2N_{S})}$  if $N_{S}=2^{k}$ and $\{\delta_{u}\}^{N_{S}}_{u=1}=e^{j\pi (6u-1)/(3N_{S})}$ if $N_{S}=3\times 2^{k}$. This design is related to the linear complex field (LCF) encoder given in [16].

For simplicity, we firstly introduce (2-source,2-relay,1-destination) system:


\subsection{(2,2,1) system}
Without losing generality, we assume that at time slot $n$, symbols from different sources are received by $R_{1}$ and $R_{2}$ simultaneously forwards the symbol received at previous TS. While at time slot $n+1$, sources' symbols are sent to $R_{2}$ and $R_{1}$ forwards the previously received symbol to $D$, as shown in Fig. 3.

{\em Time Slot n}: $S_{1}$ transmits $\theta_{1}x_{1}^{(n)}$ and simultaneously $S_{2}$ transmits $\theta_{2}x_{2}^{(n)}$ where $x_{i}^{(n)}$ denotes the modulated digital symbol sent by source $i$ at TS $n$ $(i=1,2$ and $n=1,2,...,L)$. At the same TS, $R_{2}$ sends the operated signal received at the previous TS $n-1$ to $D$. Therefore, the received signal at $R_{1}$ and $D$ are respectively given by
\begin{equation}
\begin{split}
y_{R_{1}}^{(n)} =&h_{S_{1}R_{1}}\theta_{1}x_{1}^{(n)}+h_{S_{2}R_{1}}\theta_{2}x_{2}^{(n)}\\
&+h_{12}(\theta_{1}x_{1_{r}}^{(n)}+\theta_{2}x_{2_{r}}^{(n)})+w_{R_{1}},
\end{split}
\end{equation}
\begin{equation}
y_{D}^{(n)} =h_{R_{2}D}(\theta_{1}x_{1_{r}}^{(n)}+\theta_{2}x_{2_{r}}^{(n)})+w_{D}.\quad
\end{equation}
We use subscript $r$ to denote that symbols are sent by relay nodes.
\begin{figure}[h]
\centering
\includegraphics[width=0.45\textwidth]{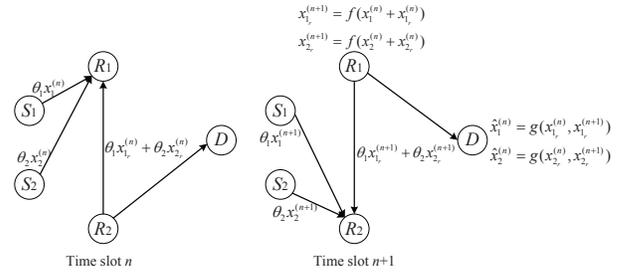}
\caption{(2,2,1) system model}
\end{figure}
Hence $\theta_{1}x_{1_{r}}^{(n)}+\theta_{2}x_{2_{r}}^{(n)}$ is the symbol sent by $R_{2}$ at TS $n$ after PNC, which will be illustrated in the next section. All the symbols $x_{i}^{(n)},x_{i_{r}}^{(n)}$ transmitted by sources and relays are drawn from a finite alphabet $\mathcal{A}$$_{x}$ with cardinality $|\mathcal{A}$$_{x}|$, which is determined by modulation mode. Maximum likelihood (ML) detection is performed at $R$ and $D$ as
\begin{equation}
\begin{split}
[(&\hat{x}_{1}^{(n)},\hat{x}_{1_{r}}^{(n)}),(\hat{x}_{2}^{(n)},\hat{x}_{2_{r}}^{(n)})]_{R}=\arg\min_{x_{i}^{(n)},x_{i_{r}}^{(n)}\in {\mathcal{A}}_x}||y_{R_{1}}^{(n)}-\\
&\theta_{1}(h_{S_{1}R_{1}}x_{1}^{(n)}+h_{12}x_{1_{r}}^{(n)})-\theta_{2}(h_{S_{2}R_{1}}x_{2}^{(n)}+ h_{12}x_{2_{r}}^{(n)})||^{2}
\label{(3)}
\end{split}
\end{equation}
\begin{equation}
\begin{split}
\label{4}
(\hat{x}_{1_{r}}^{(n)},\hat{x}_{2_{r}}^{(n)})_{D}=\arg\min_{x_{i_{r}}^{(n)}\in {\mathcal{A}}_x}\!\!\!\!||y_{D}^{(n)}- h_{R_{2}D}(\theta_{1}x_{1_{r}}^{(n)}+\theta_{2}x_{2_{r}}^{(n)})||^{2}
\end{split}
\end{equation}
where $\theta_{1}$ and $\theta_{2}$ are used to identify the symbols from $S_{1}$ and $S_{2}$ respectively. If there is no interference ($x_{1_{r}}^{(n)}$ and $x_{2_{r}}^{(n)}$), relay node is able to recover $x_{1}^{(n)}$ and $x_{2}^{(n)}$ [3]. Therefore, due to the IRI, the relay just determines $\hat{x}_{i}^{(n)}+\hat{x}_{i_{r}}^{(n)}$, rather than $(\hat{x}_{i}^{(n)},\hat{x}_{i_{r}}^{(n)})$. Then, we adopt PNC to deal with the interference. The core idea of PNC is encoding two sources' symbols at relay and decoding at destination [1]. In our scheme, the combinations of desired symbols and interference ($\hat{x}_{i}^{(n)}+\hat{x}_{i_{r}}^{(n)}$) are encoded through function $f$ at relay and separated through function $g$ at $D$. Let $f: \{x_{a}+x_{b}|x_{a},x_{b}\in{{\mathcal{A}}_x}\}\rightarrow {{\mathcal{A}}_x}$ denote the many-to-one mapping function. We set $x_{1_{r}}^{(n+1)}=f(\hat{x}_{1}^{(n)}+\hat{x}_{1_{r}}^{(n)})$ and $x_{2_{r}}^{(n+1)}=f(\hat{x}_{2}^{(n)}+\hat{x}_{2_{r}}^{(n)})$ to be the symbols for $R_{1}$ to forward within the next TS.

{\em Time Slot $n+1$}: $S_{1}$ transmits $\theta_{1}x_{1}^{(n+1)}$ and simultaneously $S_{2}$ transmits $\theta_{2}x_{2}^{(n+1)}$, which are received by $R_{2}$. At the same TS, $R_{1}$ forwards $\theta_{1}x_{1_{r}}^{(n+1)}+\theta_{2}x_{2_{r}}^{(n+1)}$ to $D$. The received signal at $D$ are given by
\begin{equation}
y_{D}^{(n+1)} =h_{R_{1}D}(\theta_{1}x_{1_{r}}^{(n+1)}+\theta_{2}x_{2_{r}}^{(n+1)})+w_{D},
\end{equation}
and the ML detector at $D$ can be expressed as
\begin{equation}
\begin{split}
\label{6}
(\hat{x}_{1_{r}}^{(n+1)},\hat{x}_{2_{r}}^{(n+1)})_{D}=&\arg\min_{x_{i_{r}}^{(n)}\in {\mathcal{A}}_x}||y_{D}^{(n+1)}\\-& h_{R_{1}D}(\theta_{1}x_{1_{r}}^{(n+1)}+\theta_{2}x_{2_{r}}^{(n+1)})||^{2}.
\end{split}
\end{equation}

With the purpose of detecting the sources symbols: $x_{1}^{(n)},x_{2}^{(n)}$, we introduce another many-to-one function $g:({{\mathcal{A}}_x},{{\mathcal{A}}_x})\rightarrow {{\mathcal{A}}_x}$. According to the symbols detected at TS $n$ and $n+1$, given by $(\ref{4})$ and $(\ref{6})$, sources symbols can be recovered by
\begin{equation}
\begin{split}
&(\hat{x}_{1}^{(n)},\hat{x}_{2}^{(n)})=(g(\hat{x}_{1_{r}}^{(n)},\hat{x}_{1_{r}}^{(n+1)}),g(\hat{x}_{2_{r}}^{(n)},\hat{x}_{2_{r}}^{(n+1)}))\label{(7)}\\
&=(g(\hat{x}_{1_{r}}^{(n)},f(\hat{x}_{1}^{(n)}+\hat{x}_{1_{r}}^{(n)})),g(\hat{x}_{2_{r}}^{(n)},f(\hat{x}_{2}^{(n)}+\hat{x}_{2_{r}}^{(n)}))).
\end{split}
\end{equation}
The original symbol $x_{i}^{(n)}$ is recovered through $g(\hat{x}_{i_{r}}^{(n)},f(\hat{x}_{i}^{(n)},$ $\hat{x}_{i_{r}}^{(n)}))$. Depending on the $y_{D}^{(n+1)}$ and $y_{D}^{(n)}$ stored at $D$, all the sources' symbols can be recovered through $(\ref{(7)})$. The key issue is that whether the function $f$ and $g$ are able to successfully cancel the inter-relay interference (IRI), which will be detailed in the next section.

\subsection{($N_{S}$,2,1) system}
The transmission scheme can be generalized to the multi-source ($N_{S}$,2,1) networks where $N_{S}\geq2$.

{\em Time Slot n}: According to the previous definition of $\mathbf{x}$ and $\Theta^{T}$, the I/O relationships are
\begin{equation}
\begin{split}
y_{R_{1}}^{(n)}=\Theta^{T}{\mathbf{H}}_{SR}{\mathbf{x}}^{(n)}+h_{12}\Theta^{T}{\mathbf{x}}^{(n)}_{r}+w_{R_{1}},
\end{split}
\end{equation}
\begin{equation}
y_{D}^{(n)} =h_{R_{2}D}\Theta^{T}{\mathbf{x}}^{(n)}_{r}+w_{D},\quad\quad\quad\quad\quad\quad\:
\end{equation}
where ${\mathbf{H}}_{SR}=diag(h_{S_{1}R_{1}},h_{S_{2}R_{1}},...,h_{S_{N_{S}}R_{1}})$. ${\mathbf{x}}^{(n)}_{r}=[x_{1_{r}}^{(n)},x_{2_{r}}^{(n)},...,x_{N_{S_{r}}}^{(n)}]^{T}$ denotes the symbols sent by relay at TS $n$ before multiplied by $\Theta^{T}$. Through ML detection at relay, the symbol required to be forwarded by relay within the next TS is generated by ${\mathbf{x}}^{(n+1)}_{r}=[f(\hat{x}_{1}^{(n)}+\hat{x}_{1_{r}}^{(n)}),f(\hat{x}_{2}^{(n)}+\hat{x}_{2_{r}}^{(n)}),$ $...,f(\hat{x}_{N_{S}}^{(n)}+\hat{x}^{(n)}_{N_{S_{r}}})]^{T}$.

{\em Time Slot $n+1$}: $N_{S}$ sources continue broadcast symbols and simultaneously $R_{1}$ forwards $\Theta^{T}{\mathbf{x}}^{(n+1)}_{r}$ to $D$ [17]. The received signal at $D$ is given by
\begin{equation}
y_{D}^{(n+1)} =h_{R_{1}D}\Theta^{T}{\mathbf{x}}^{(n+1)}_{r}+w_{D},
\end{equation}
after which ${\mathbf{x}}^{(n+1)}_{r}$ can be obtained at $D$ by:
\begin{equation}
\begin{split}
\hat{{\mathbf{x}}}^{(n+1)}_{r}=\arg\min_{x_{i_{r}}^{(n+1)}\in {\mathcal{A}}_x}||y_{D}^{(n+1)}- h_{R_{1}D}\Theta^{T}{\mathbf{x}}^{(n+1)}_{r}||^{2}.
\end{split}
\end{equation}

Therefore, having $\hat{{\mathbf{x}}}^{(n+1)}_{r}=[\hat{x}^{(n+1)}_{1_{r}},\hat{x}^{(n+1)}_{2_{r}},...,\hat{x}^{(n+1)}_{N_{S_{r}}}]^{T}$, together with $\hat{{\mathbf{x}}}^{(n)}_{r}$ that estimated at last TS, each original symbol $x_{i}^{(n)}$ in ${\mathbf{x}}^{(n)}=[x_{1}^{(n)},x_{2}^{(n)},...,x_{N_{S}}^{(n)}]^{T}$ is recovered through
\begin{equation}
\begin{split}
\label{(12)}
\hat{x}_{i}^{(n)}=&g(\hat{x}_{i_{r}}^{(n)},\hat{x}_{i_{r}}^{(n+1)})\\
=&g(\hat{x}_{i_{r}}^{(n)},f(\hat{x}_{i}^{(n)}+\hat{x}_{i_{r}}^{(n)}))\quad (i=1,2,...,N_{S}).
\end{split}
\end{equation}

Finally, $N_{S}$ sources successfully send $L\times N_{S}$ symbols during $L+1$ TSs. As long as the destination stores current and former estimated symbols $(\hat{\mathbf{x}}^{(n+1)}_{r},\hat{\mathbf{x}}^{(n)}_{r})$, sources symbols ${\mathbf{x}}^{(n)}$ can be recovered through $(\ref{(12)})$. Such transmission scheme achieves throughput $L/(L+1)$ sym/S/TS after the IRI cancellation (IRIC). When $L$ is large enough, the ideal throughput reaches 1 sym/S/TS.

\section{IRI Cancellation with PNC}
Based on the discussion in the previous section, the symbols received by relay from sources are interfered with the signal from the other relay. For the fact that relays are unable to extract the desired symbols of sources, we introduce PNC to deal with the interference. We use function $f$ at relay nodes and function $g$ at destination to achieve IRI cancellation (IRIC). Since $\hat{x}_{i}^{(n)}=g(\hat{x}_{i_{r}}^{(n)},f(\hat{x}_{i}^{(n)}+\hat{x}_{i_{r}}^{(n)}))$, the most straightforward method is making use of the fact that $b\oplus a\oplus b=a$ where $\oplus$ denotes bitwise exclusive OR (XOR) operation. Hence $\hat{x}_{i}^{(n)}$ is recovered through
\begin{equation}
\begin{split}
\label{(13)}
g(\hat{x}_{i_{r}}^{(n)},f(\hat{x}_{i}^{(n)}+\hat{x}_{i_{r}}^{(n)}))=&\hat{x}_{i_{r}}^{(n)}\oplus f(\hat{x}_{i}^{(n)}+\hat{x}_{i_{r}}^{(n)})\\=&\hat{x}_{i_{r}}^{(n)}\oplus (\hat{x}_{i}^{(n)}\oplus\hat{x}_{i_{r}}^{(n)})=\hat{x}_{i}^{(n)}.
\end{split}
\end{equation}


To clearly illustrate the IRIC scheme, we make it simplify that $h_{link}\equiv 1$. Without losing generality, all sources and relays adopt the same modulation mode: $x_{i}^{(n)},x_{i_{r}}^{(n+1)}\in {{\mathcal{A}}_x} \, (n=1,2,...,L)$. However, the symbols transmitted are $\Theta^{T}{\mathbf{x}}=s$, rather than $\mathbf{x}$. Satisfying the request for one-to-one mapping between $\mathbf{x}$ and $s$, $N_{S}$ sources generate $M^{N_{S}}$ points in the constellation of $s$ where $M=|{\mathcal{A}}_{x}|$. Linear growth of $N_{S}$ leads to exponential growth of cardinality of $s$. Therefore, we mainly consider modulation modes with low $M$ such as BPSK$(M=2)$ and QPSK$(M=4)$, in order to achieve better anti-noise performance in multi-source system.

\begin{table*}[ht]
\centering
\small
\caption{IRIC for BPSK}
\renewcommand{\arraystretch}{1.5}
\begin{tabular}[ht]{|c|c|c||c|c|c|}

\hline
\multicolumn{3}{|c||}{TS $n$} & \multicolumn{3}{c|}{TS $n+1$}\\ \cline{1-6}
$y_{D}^{(n)}$ & Sources' symbols & Received by $R$ with IRI & $f(\hat{x}^{(n)}_{i_{r}}+\hat{x}^{(n)}_{i})$ & $y_{D}^{(n+1)}$ & $g({\hat{x}}^{(n)}_{i_{r}},\hat{x}^{(n+1)}_{i_{r}})$ \\ \cline{1-6}
${\mathbf{\hat{x}}}^{(n)}_{r}$ & ${\mathbf{x}}^{(n)}$ & ${\mathbf{x}}^{(n)}_{r}+{\mathbf{x}}^{(n)}$ & ${\mathbf{x}}^{(n+1)}_{r}$ & ${\mathbf{\hat{x}}}^{(n+1)}_{r}$ & ${\mathbf{\hat{x}}}^{(n)}$\\ \cline{1-6}
(1,-1)&(1,1)&(2,0)&(-1,1)&(-1,1)&(1,1)\\ \cline{1-6}
(1,-1)&(-1,-1)&(0,-2)&(1,-1)&(1,-1)&(-1,-1)\\ \cline{1-6}
\hline
\end{tabular}
\end{table*}

\begin{table*}[ht]
\centering
\small
\renewcommand{\arraystretch}{1.5}
\caption{IRIC for QPSK}
\begin{tabular}[ht]{|c|c|c||c|c|c|}
\hline
\multicolumn{3}{|c||}{TS $n$} & \multicolumn{3}{c|}{TS $n+1$}\\ \cline{1-6}
$y_{D}^{(n)}$ & Sources' symbols & Received by $R$ with IRI & $f(\hat{x}^{(n)}_{i_{r}}+\hat{x}^{(n)}_{i})$ & $y_{D}^{(n+1)}$ & $g({\hat{x}}^{(n)}_{i_{r}},\hat{x}^{(n+1)}_{i_{r}})$ \\ \cline{1-6}
${\mathbf{\hat{x}}}^{(n)}_{r}$ & ${\mathbf{x}}^{(n)}$ & ${\mathbf{x}}^{(n)}_{r}+{\mathbf{x}}^{(n)}$ & ${\mathbf{x}}^{(n+1)}_{r}$ & ${\mathbf{\hat{x}}}^{(n+1)}_{r}$ & ${\mathbf{\hat{x}}}^{(n)}$\\ \cline{1-6}
(1+$j$,1-$j$)&(1+$j$,1+$j$)&(2+2$j$,2)&(-1-$j$,-1+$j$)&(-1-$j$,-1+$j$)&(1+$j$,1+$j$)\\ \cline{1-6}
(1+$j$,1-$j$)&(-1+$j$,-1-$j$)&(2$j$,-2$j$)&(1-$j$,1-$j$)&(1-$j$,1-$j$)&(-1+$j$,-1-$j$)\\ \cline{1-6}
(1+$j$,1-$j$)&(-1-$j$,1-$j$)&(0,2-2$j$)&(1+$j$,-1-$j$)&(1+$j$,-1-$j$)&(-1-$j$,1-$j$)\\ \cline{1-6}
(-1-$j$,1-$j$)&(-1-$j$,-1+$j$)&(-2-2$j$,0)&(-1-$j$,1+$j$)&(-1-$j$,1+$j$)&(-1-$j$,-1+$j$)\\ \cline{1-6}
(-1-$j$,-1+$j$)&(-1+$j$,-1+$j$)&(-2,-2+2$j$)&(-1+$j$,-1-$j$)&(-1+$j$,-1-$j$)&(-1+$j$,-1+$j$)\\ \cline{1-6}
\hline
\end{tabular}
\end{table*}

\subsection{BPSK modulation}
BPSK modulation generates two points in the constellation of each original symbol $x$. Regardless of high frequency carrier, the baseband digital symbol can be expressed as ${\mathcal{A}}_{x}=\{1,-1\}$ and $M=|{\mathcal{A}}_{x}|=2$. Under the assumption that $h_{link}\equiv 1$, the symbol detected by relay at TS $n$ is $x_{i}^{(n)}+x_{i_{r}}^{(n)}\in \{2,0,-2\}$. The estimated symbol is given by
\begin{equation}
\begin{split}
\label{(14)}
(\hat{x}_{1}^{(n)}+&\hat{x}_{1_{r}}^{(n)},\hat{x}_{2}^{(n)}+\hat{x}_{2_{r}}^{(n)},...,\hat{x}_{N_{S}}^{(n)}+\hat{x}_{N_{S_{r}}}^{(n)})_{R}\\=&\arg\min_{x_{i}^{(n)}+x_{i_{r}}^{(n)}\in {\mathcal{A}}_{z}}||y_{R_{1}}^{(n)}-\Theta^{T}({\mathbf{x}}^{(n)}+{\mathbf{x}}^{(n)}_{r})||^{2}
\end{split}
\end{equation}
where ${\mathcal{A}}_{z}$ denotes $\{z=x_{i}^{(n)}+x_{i_{r}}^{(n)}|x_{i}^{(n)},x_{i_{r}}^{(n)}\in {\mathcal{A}}_{x}\}$. Hence we can express the function $f$ as
\begin{equation}
f(\hat{x}_{i}^{(n)}+\hat{x}_{i_{r}}^{(n)})=\hat{x}_{i}^{(n)}\oplus\hat{x}_{i_{r}}^{(n)}=1-|\hat{x}_{i}^{(n)}+\hat{x}_{i_{r}}^{(n)}|,
\end{equation}
from which $f(2)=f(-2)=-1$ and $f(0)=1$. As long as satisfying $a=b\oplus a\oplus b$, note that the so-called $\oplus$ not only means $1\oplus 1=0, 1\oplus0=1$, but also varies based on different I/O relationships.

At TS $n+1$, $\Theta^{T}{\mathbf{x}}^{(n+1)}_{r}=\Theta^{T}[f(\hat{x}_{1}^{(n)}+\hat{x}_{1_{r}}^{(n)}),f(\hat{x}_{2}^{(n)}+\hat{x}_{2_{r}}^{(n)}),...,f(\hat{x}_{N_{S}}^{(n)}+\hat{x}_{N_{S_{r}}}^{(n)})]^{T}$ is relayed to $D$. After detecting ${\mathbf{\hat{x}}}^{(n+1)}_{r}$, together with ${\mathbf{\hat{x}}}^{(n)}_{r}$ stored at TS $n$, another XOR operation is performed through $(\ref{(13)})$. In order to recover $x_{i}^{(n)}$, $g(1,1)=g(-1,-1)=-1$ and $g(1,-1)=g(1,-1)=1$ are supposed to be satisfied. Therefore, function $g$ can be defined as
\begin{equation}
\begin{split}
g(\hat{x}_{i_{r}}^{(n)},f(\hat{x}_{i}^{(n)}+\hat{x}_{i_{r}}^{(n)}))=1-|\hat{x}_{i_{r}}^{(n)}+f(\hat{x}_{i}^{(n)}+\hat{x}_{i_{r}}^{(n)})|.
\end{split}
\end{equation}

It is clear from $(\ref{(13)})$ that IRIC is completely successful and original symbols from sources are recovered at destination. The difference between multi-source system and single-source system is the dimensions of vector $\mathbf{x}$, among which pairwise dimensional components are relative independent and separated by $\Theta^{T}_{N_{S}\times 1}$ corresponding to the $N_{S}$ sources. Therefore, without losing generality, taking 2-source system as an example, a complete transmission process is presented in Table II. The symbols in the last column are totally the same as the original symbols listed in the second column. Therefore, IRIC is completely achieved at destination.

\subsection{QPSK modulation}
QPSK modulation: ${\mathcal{A}}_{x}=\{1+j,-1+j,1-j,-1-j\}$ with $M=|{\mathcal{A}}_{x}|=4$. QPSK symbols can be considered as the combination of two BPSK symbols: the in-phase and the quadrature-phase components. Estimated symbols $(\hat{x}_{1}^{(n)}+\hat{x}_{1_{r}}^{(n)},\hat{x}_{2}^{(n)}+\hat{x}_{2_{r}}^{(n)},...,\hat{x}_{N_{S}}^{(n)}+\hat{x}_{N_{S_{r}}}^{(n)})_{R}$ are obtained through $(\ref{(14)})$ where ${\mathcal{A}}_{z}=\{2,-2,0,2j,-2j,2+2j,2-2j,-2+2j,-2-2j\}$. Although $\hat{x}_{i}^{(n)}$ and $\hat{x}_{i_{r}}^{(n)}$ are not separable, the in-phase component $\Re\{\hat{x}_{i}^{(n)}+\hat{x}_{i_{r}}^{(n)}\}$ and the quadrature-phase component $\Im\{\hat{x}_{i}^{(n)}+\hat{x}_{i_{r}}^{(n)}\}$ can be easily separated. Therefore, it is not necessary to change the definition of $f$ and $g$ employed in BPSK modulation. The PNC operation at relay can be expressed as
\begin{equation}
x_{i_{r}}^{(n+1)}=f(\Re\{\hat{x}_{i}^{(n)}+\hat{x}_{i_{r}}^{(n)}\})+j\times f(\Im\{\hat{x}_{i}^{(n)}+\hat{x}_{i_{r}}^{(n)}\}).
\end{equation}

At TS $n+1$, after detection of ${\mathbf{\hat{x}}}^{(n+1)}_{r}$, together with ${\mathbf{\hat{x}}}^{(n)}_{r}$ stored at TS $n$, another XOR operation on divided in-phase and quadrature-phase components of the received symbols is performed through
\begin{equation}
\begin{split}
\hat{x}_{i}^{(n)}=&g(\Re\{\hat{x}_{i_{r}}^{(n)}\},f(\Re\{\hat{x}_{i}^{(n)}+\hat{x}_{i_{r}}^{(n)}\}))\\
&+j\times g(\Im\{\hat{x}_{i_{r}}^{(n)}\},f(\Im\{\hat{x}_{i}^{(n)}+\hat{x}_{i_{r}}^{(n)}\})).
\end{split}
\end{equation}

Table III shows the transmission process of 2-source system with QPSK modulation. Similar to that of BPSK modulation, applying the same $f$ and $g$ at relay and destination respectively, the symbols from sources are completely recovered and entire transmission process is IRI free.

\subsection{General PNC process}
In practice, higher order modulation modes might be used, such as 16-QAM, 64-QAM [18] and so on. Apart from BPSK and QPSK modulation, our scheme can be extended to other modulation methods with different definition of $f$ and $g$ to achieve PNC mapping. To summarize, a general PNC process consists of the following steps:\\
1) At each relay, for each source $S_{i}$, many-to-one modulation mapping, $f$: $(x^{(n)}_{i}+x^{(n)}_{i_{r}})\rightarrow x^{(n+1)}_{i_{r}}$.\\
2) Form the signal for relay to send: $\theta_{1}x^{(n+1)}_{1_{r}}+\theta_{2}x^{(n+1)}_{2_{r}}+\cdots+\theta_{N_{S}}x^{(n+1)}_{N_{S_{r}}}$.\\
3) At $D$, for each source $S_{i}$, one-to-one demodulation mapping, $g$: $(x^{(n+1)}_{i_{r}},x^{(n)}_{i_{r}})\rightarrow x^{(n)}_{i}$.\\
The designs of $f$ and $g$ vary based on different modulation modes, which are detailed in [1].

Now we consider a more practical system under flat slow fading channels. We assume that CSI is available at receiver nodes. At TS $n$, relay node $R_{1}$ is aware of the CSI of link $S\text{\bfseries{--}}R$:$(h_{S_{1}R_{1}},h_{S_{2}R_{1}},...,h_{S_{N_{S}}R_{1}})$ and the CSI of link $R\text{\bfseries{--}}R$:$h_{12}$. Hence, $(\ref{(3)})$ can be generalized to $N_{S}$-sources system, expressed as
\begin{equation}
\begin{split}
[(&\hat{x}_{1}^{(n)},\hat{x}_{1_{r}}^{(n)}),(\hat{x}_{2}^{(n)},\hat{x}_{2_{r}}^{(n)}),...,(\hat{x}_{N_{S}}^{(n)},\hat{x}_{N_{S_{r}}}^{(n)})]_{R}=\\
&\arg\min_{x_{i}^{(n)}\in {\mathcal{A}}_x}||y_{R_{1}}^{(n)}-\Theta^{T}({\mathbf{H}}_{SR}{\mathbf{x}}^{(n)}+h_{12}{\mathbf{x}}^{(n)}_{r})||^{2}.
\end{split}
\end{equation}
We still mainly consider $\hat{x}_{i}^{(n)}+\hat{x}_{i_{r}}^{(n)}$. The symbol corresponding to source $i$ for relay to forward can be obtained through $f(\hat{x}_{i}^{(n)}+\hat{x}_{i_{r}}^{(n)})$. The IRIC at $D$ is the same as the algorithm discussed above. In general, the total transmission process is similar to that in the system free from channel fading.


\section{Lower Bound for System Throughput}
In this section, we mainly analyze the throughput performance for the $(N_{S},2,1)$ system as described above. With the purpose of giving proof to verify that the proposed transmission scheme with IRIC can achieve 1 sym/S/TS under ideal circumstance, we assume that $h_{link}\equiv 1$. All the additive complex White Gaussian noise (AWGN) $w\sim {\mathcal{CN}}(0,N_{0})$. Both sources and relays adopt the same modulation mode, i.e., ${\mathcal{A}}_{x}$ with $M=|{\mathcal{A}}_{x}|$.

\subsection{SEP of $R\text{\bfseries{--}}D$ link}
Firstly, we consider the SEP performance of $R\text{\bfseries{--}}D$ link. In traditional transmission procedure with single-source, the SEP of source symbol $x_{a} (x_{a}\in{\mathcal{A}}_{x})$ can be calculated as
\begin{equation}
\begin{split}
\label{20}
P_{e}\{x_{a}\}=\sum_{\begin{subarray}{c}x_{b}\in{\mathcal{A}}_{x} \\ x_{b}\neq x_{a}\end{subarray}}P\{x_{a}\rightarrow x_{b}\},
\end{split}
\end{equation}
where $P\{x_{a}\rightarrow x_{b}\}$ denotes the pairwise error probability of mistaking $x_{b}$ for $x_{a}$. However, deploying CFNC in multi-source system, the transmitted symbol is $\Theta^{T}{\mathbf{x}}=s$, which contains the information of all sources.
$N_{S}$ sources generate $M^{N_{S}}$ points in the constellation of $s$. However, the incorrect detection of $s$ does not represent the error of all sources' symbols $[x_{1},x_{2},...,x_{N_{S}}]$. As an example to explain the idea, we assume $N_{S}=2$ and BPSK $(M=|{\mathcal{A}}_{x}|=2)$ modulation is adopted by senders. Let $\theta_{1}=1$ and $\theta_{2}=j$, as illustrated in Fig. 4. $M^{N_{S}}=4$ points exist in the constellation of $s$. Without losing generality, we assume the sources symbols are $[-1,1]$, and thus $s=\theta_{1}\times (-1)+\theta_{2}\times 1$, marked in the constellation. Even if the detected $\hat{s}\neq s$, for example $\hat{s}=\theta_{1}\times 1+\theta_{2}\times 1$, equivalent to $\hat{\mathbf{x}}=[1,1]$, from which one of the two sources' symbol is successfully recovered. Let random variable $\xi$ denote the number of symbols failed to be recovered among $N_{S}$ sources. Therefore, the probability distribution of $\xi\, (0\leq\xi\leq N_{S})$ is given in Table IV where $P_{e}$ denotes the SEP of $s_{a}$. The probability $P_{e-k}\:(k=1,2,...,N_{S})$, denoting $k$ of $N_{S}$ symbols failed to be recovered at receiver, is given by
\begin{equation}
\begin{split}
\label{21}
P_{e-k}=\sum_{s_{b}\in{\mathcal{A}}_{s}}P\{s_{a}\rightarrow s_{b}\,|\:||{\mathbf{x}}_{a}-{\mathbf{x}}_{b}||_{0}=k\},
\end{split}
\end{equation}
where $s_{a}=\Theta^{T}{\mathbf{x}}_{a},s_{b}=\Theta^{T}{\mathbf{x}}_{b}$. $||{\mathbf{x}}_{a}-{\mathbf{x}}_{b}||_{0}$ indicates the zero-norm which denotes the number of nonzero elements of ${\mathbf{x}}_{a}-{\mathbf{x}}_{b}$.
Hence, the SEP of $R\text{\bfseries{--}}D$ link for each source can be expressed as
\begin{equation}
\begin{split}
\label{22}
P_{e-RD}=E\{\xi\}/N_{S}=\frac{1}{N_{S}}\sum_{k=1}^{N_{S}}k\times P_{e-k},
\end{split}
\end{equation}
where $E\{\xi\}$ is the mathematical expectation of $\xi$. For any given constellation, the SEP in AWGN channel can be estimated by union bound[19]:
\begin{equation}
\label{23}
P_{e}(s_{a})=\sum_{\begin{subarray}{c}b=1 \\ a\neq b\end{subarray}}^{|{\mathcal{A}}_{s}|}P\{s_{a}\rightarrow s_{b}\}\leq \sum_{\begin{subarray}{c}b=1 \\ a\neq b\end{subarray}}^{|{\mathcal{A}}_{s}|}Q\left(\frac{d_{ab}^{(s)}}{\sqrt{2N_{0}}}\right),
\end{equation}
where $N_{0}$ is the noise power spectral density of the channel and $d_{ab}^{(s)}$ denotes the Euclidean distance between $s_{a}$ and $s_{b}$ in the constellation.
\begin{figure}[h]
\centering
\includegraphics[width=0.45\textwidth]{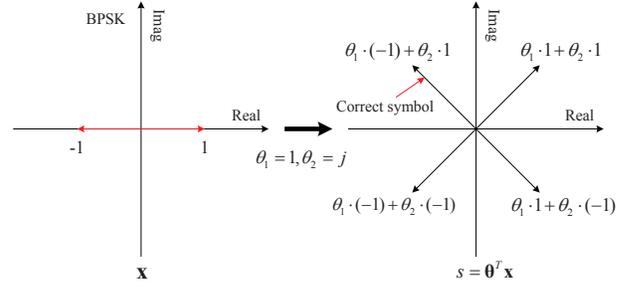}
\caption{The constellations of symbols before and after CFNC}
\label{f4}
\end{figure}
\begin{table}[ht]
\centering
\caption{Probability distribution of $\xi$}
\renewcommand{\arraystretch}{1.5}
\begin{tabular}[ht]{|c||c|c|c|c|c|c|}

\hline

$\xi$: & 0 & 1 & 2 & $\cdots$ & $N_{S}-1$ & $N_{S}$ \\ \cline{1-7}
P: & $1-P_{e}$ & $P_{e-1}$ & $P_{e-2}$ & $\cdots$ & $P_{e-(N_{S}-1)}$ & $P_{e-N_{S}}$\\ \cline{1-7}
\hline
\end{tabular}
\end{table}

We assume the uniform distribution of $s$ due to one-to-one mapping between $s$ and $\mathbf{x}$, i.e. $P\{s_{a}\}=1/|{\mathcal{A}}_{s}|\:(a=1,2,...,|{\mathcal{A}}_{s}|)$. Based on $(\ref{23})$, $P_{e-k}$ is given by
\begin{equation}
\label{24}
P_{e-k}\leq \frac{1}{|{\mathcal{A}}_{s}|}\sum_{a=1}^{|{\mathcal{A}}_{s}|}\left(\sum_{\begin{subarray}{c}b=1,\\
||{\mathbf{x}}_{a}-{\mathbf{x}}_{b}||_{0}=k\end{subarray}}^{|{\mathcal{A}}_{s}|}Q\left(\frac{d_{ab}^{(s)}}{\sqrt{2N_{0}}}\right)\right),
\end{equation}
according to which $(\ref{22})$ can be rewritten as
\begin{equation}
\begin{split}
\label{25}
P_{e-RD}\leq\frac{1}{N_{S}|{\mathcal{A}}_{s}|}\sum_{k=1}^{N_{S}}k \sum_{a=1}^{|{\mathcal{A}}_{s}|}\left(\sum_{\begin{subarray}{c}b=1,\\||{\mathbf{x}}_{a}-{\mathbf{x}}_{b}||_{0}=k
\end{subarray}}^{|{\mathcal{A}}_{s}|}Q\left(\frac{d_{ab}^{(s)}}{\sqrt{2N_{0}}}\right)\right).
\end{split}
\end{equation}

\subsection{SEP of $S\text{\bfseries{--}}R$ link}
Secondly, we consider the SEP of $S\text{\bfseries{--}}R$ link. Without losing generality, we set $N_{0-SR}=N_{0-RD}=N_{0}$. Links are free from channel fading and the signal received at $R$ is the superposition of each node's symbol. The transmissions from $N_{S}$ sources and one of the relays to the other relay can be treated as (1-source,1-relay) link where the symbol sent by the equivalent source is $\Theta^{T}({\mathbf{x}}+{\mathbf{x}}_{r})$. Therefore, the analysis for $R\text{\bfseries{--}}D$ link can be applied to the derivation of SEP for $S\text{\bfseries{--}}R$ link. However, one of the differences between $S\text{\bfseries{--}}R$ and $R\text{\bfseries{--}}D$ link is that the symbol is $\Theta^{T}({\mathbf{x}}+{\mathbf{x}}_{r})=\Theta^{T}\mathbf{z}$, rather than $\Theta^{T}\mathbf{x}$. Let $y=\Theta^{T}\mathbf{z}$. The most significant difference between $s$ and $y$ is that any $s_{a}\in{\mathcal{A}}_{s}$ appears with equal prior probability due to one-to-one mapping from $\mathbf{x}$ on the condition that each symbol in ${\mathcal{A}}_{x}$ has the same prior probability. However, the prior probability of $y_{a}\in{\mathcal{A}}_{y}$ varies due to many-to-one mapping $({\mathbf{x}},{\mathbf{x}}_{r})\rightarrow ({\mathbf{x}}+{\mathbf{x}}_{r})$. For instance, deploying BPSK modulation, ${\mathcal{A}}_{z}=\{-2,0,2\}$ with corresponding probability of 0.25, 0.5, 0.25, respectively. Therefore, when calculating $P_{e-k}$, $(\ref{24})$ can be rewritten as
\begin{equation}
\label{26}
P_{e-k}\leq \sum_{a=1}^{|{\mathcal{A}}_{y}|}P\{y_{a}\}\left(\sum_{\begin{subarray}{c}b=1,\\
||{\mathbf{\mathbf{z}}}_{a}-{\mathbf{\mathbf{z}}}_{b}||_{0}=k\end{subarray}}^{|{\mathcal{A}}_{y}|}Q\left(\frac{d_{ab}^{(y)}}{\sqrt{2N_{0}}}\right)\right),
\end{equation}
where $P\{y_{a}\}$ is the priori probability of $y_{a}$ and $d_{ab}^{(y)}$ denotes the Euclidean distance between $y_{a}$ and $y_{b}$. Hence the SEP of $S\text{\bfseries{--}}R$ link is given by
\begin{equation}
\begin{split}
\label{27}
P_{e-SR}\leq\frac{1}{N_{S}}\sum_{k=1}^{N_{S}}k \sum_{a=1}^{|{\mathcal{A}}_{y}|}P\{y_{a}\}\left(\sum_{\begin{subarray}{c}b=1,\\||{\mathbf{z}}_{a}-{\mathbf{z}}_{b}||_{0}=k
\end{subarray}}^{|{\mathcal{A}}_{y}|}Q\left(\frac{d_{ab}^{(y)}}{\sqrt{2N_{0}}}\right)\right).
\end{split}
\end{equation}
\subsection{Lower bound for throughput with CFNC}
Traditional CFNC [3] requires $2L$ TSs to transmit $L$ continuous symbols: at odd TSs, sources send symbols to relays; at even TSs, relay nodes forward the symbol to $D$ while sources keep silence.
Since we have $P_{e-SR}$ and $P_{e-RD}$, the throughput of $S\text{\bfseries{--}}R\text{\bfseries{--}}D$ with traditional CFNC can be estimated by
\begin{equation}
\label{28}
T_{CFNC}=\frac{1}{2}[(1-P'_{e-SR})(1-P_{e-RD})+P']\quad\text{(sym/S/TS)},
\end{equation}
where throughput $T$ is defined as the number of successfully transmitted symbols per source per TS (sym/S/TS) within one transmission cycle. When calculating $P_{e-RD}$, $N_{0}$ in $(\ref{25})$ is supposed to be the value corresponding to the SNR after maximal ratio combining (MRC) of ${\mathrm{SNR}}_{R_{1}D}$ and ${\mathrm{SNR}}_{R_{2}D}$. Due to no IRI in traditional cooperative communication with CFNC, $P'_{e-SR}$ can be obtained through $(\ref{25})$, following the analysis of link $R\text{\bfseries{--}}D$, rather than $(\ref{27})$. The factor $\frac{1}{2}$ means that two TSs are required to send one symbol from source to destination. The first term $(1-P'_{e-SR})(1-P_{e-RD})$  represents no incorrect decision at both $R$ and $D$, i.e. $x_{a}\stackrel{SR}{\rightarrow}x_{a}\stackrel{RD}{\rightarrow}x_{a}$, while the second term $P'$ denotes the probability that one symbol is successfully recovered at $D$ while incorrect decisions occur within the transmission, i.e. the case $x_{a}\stackrel{SR}{\rightarrow}x_{b}\stackrel{RD}{\rightarrow}x_{a}\:(x_{a}\neq x_{b})$. Therefore, $P'$ is given by
\begin{equation}
\begin{split}
\label{29}
P'=&\frac{1}{N_{S}}\sum_{N=1}^{N_{S}}\frac{1}{|{\mathcal{A}}_{s}|} \sum_{a=1}^{|{\mathcal{A}}_{s}|}\sum_{s_{b}\in{\mathcal{A}}_{a,N}}P\{s_{a}\stackrel{SR}{\rightarrow}s_{b}\}\cdot P\{s_{b}\stackrel{RD}{\rightarrow}s_{a}\}\\
\leq&\frac{1}{N_{S}|{\mathcal{A}}_{s}|}\sum_{N=1}^{N_{S}} \sum_{a=1}^{|{\mathcal{A}}_{s}|}\sum_{s_{b}\in{\mathcal{A}}_{a,N}}Q\left(\frac{d_{ab}^{(s)}}{\sqrt{2N_{0}}}\right)Q\left(\frac{d_{ba}^{(s)}}{\sqrt{2N_{0}}}\right),
\end{split}
\end{equation}
where ${\mathcal{A}}_{a,N}=\{s_{b}|s_{b}\in {\mathcal{A}}_{s}, ({\mathbf{x}}_{a})_{N}\neq ({\mathbf{x}}_{b})_{N}\}$ and $({{\mathbf{x}}_{a}})_{N}$ denotes the $N$th element of ${\mathbf{x}}_{a}=[x_{a1},x_{a2},\ldots,x_{aN_{S}}]^{T}$ $(s_{a}=\Theta^{T}{\mathbf{x}}_{a})$.

Under the assumption that $N_{0-SR}=N_{0-RD}=N_{0}$, we have $P_{e-RD}=P'_{e-SR}$ since no IRI in traditional CFNC-based scheme, which results in $T_{CFNC}=\frac{1}{2}[(1-P_{e-RD})^{2}+P']$. Hence, we achieve the lower bound for throughput (see Appendix for the proof) according to $(\ref{25})$ and $(\ref{29})$:
\begin{equation}
\begin{split}
\label{30}
&T_{CFNC}\geq\\&\frac{1}{2}[(1-\frac{1}{N_{S}|{\mathcal{A}}_{s}|}\sum_{k=1}^{N_{S}}k \sum_{a=1}^{|{\mathcal{A}}_{s}|}\left(\sum_{\begin{subarray}{c}b=1,\\||{\mathbf{x}}_{a}-{\mathbf{x}}_{b}||_{0}=k
\end{subarray}}^{|{\mathcal{A}}_{s}|}Q\left(\frac{d_{ab}^{(s)}}{\sqrt{2N_{0}}}\right)\right))^{2}\\
&+\frac{1}{N_{S}|{\mathcal{A}}_{s}|}\sum_{N=1}^{N_{S}} \sum_{a=1}^{|{\mathcal{A}}_{s}|}\sum_{s_{b}\in{\mathcal{A}}_{a,N}}Q\left(\frac{d_{ab}^{(s)}}{\sqrt{2N_{0}}}\right)Q\left(\frac{d_{ba}^{(s)}}{\sqrt{2N_{0}}}\right)].
\end{split}
\end{equation}

\subsection{Lower bound for throughput with proposed scheme}
Then, we consider the throughput of the proposed transmission scheme with IRIC. According to $(\ref{(12)})$, the detection of original symbols depends on current signal received at $D$ and the signal arrived at former TS as well. Besides, at the first TS, there is no IRI at $R_{1}$ since $R_{2}$ keeps silence, which is equivalent to traditional CFNC scheme. Hence, the SEP of the symbols transmitted at TS $n$ $(1\leq n\leq L)$ can be expressed as
\begin{equation}
\label{31}
P_{new}^{(n)}\!=\!\left\{ \begin{array}{ll}
1-(1-P'_{e-SR})(1-P_{e-RD})-P' &n=1\\
1-(1-P_{e-SR})(1-P_{e-RD})^{2}-P'' &n\in[2,L]
\end{array}\right .
\end{equation}
$L+1$ TSs are required to transmit $L$ symbols for each source. Therefore, the throughput of proposed transmission scheme with IRIC is given by
\begin{equation}
\label{32}
T_{new}=\frac{1}{L+1}\sum_{n=1}^{L}(1-P_{new}^{(n)})\quad\text{(sym/S/TS)}.
\end{equation}
When $L \to \infty$, we have
\begin{equation}
\begin{split}
\label{33}
T_{new}= &{\mathop {\lim }\limits_{L \to \infty }}\frac{1}{L+1}\sum_{n=1}^{L}(1-P_{new}^{(n)})\\=&(1-P_{e-SR})(1-P_{e-RD})^{2}+P'',
\end{split}
\end{equation}
where the first term $(1-P_{e-SR})(1-P_{e-RD})^{2}$ indicates that the respective symbol is successfully transmitted from sender to receiver without errors in each link: $R\stackrel{\mathrm{TS}\: n}{\longrightarrow}D$, $S\stackrel{\mathrm{TS}\: n}{\longrightarrow}R$ and $R\stackrel{\mathrm{TS}\: n+1}{\longrightarrow}D$. $P''$ denotes the probability that incorrect decisions occur at least twice among all three links while the source's symbol still can be recovered by $D$. The calculation of $P''$ is divided into two cases:
\subsubsection{case 1}
No error occurs in $R\stackrel{\mathrm{TS}\: n}{\longrightarrow}D$, i.e. $x_{a}\stackrel{SR}{\rightarrow}x_{b}\stackrel{RD}{\rightarrow}x_{a}\:(x_{a}\neq x_{b})$ at TS $n+1$. Similar to the analysis on traditional CFNC, the probability of case 1 can be expressed as
\begin{equation}
\begin{split}
\label{34}
P_{case-1}=(1-P_{e-RD})\cdot P'''.
\end{split}
\end{equation}
Similar to $(\ref{29})$, $P'''$ is
\begin{equation}
\begin{split}
\label{35}
P'''=&\frac{1}{N_{S}}\sum_{N=1}^{N_{S}} \sum_{a=1}^{|{\mathcal{A}}_{y}|}P\{y_{a}\}\sum_{y_{b}\in{\mathcal{A}}_{a,N}}P\{y_{a}\stackrel{SR}{\rightarrow}y_{b}\} P\{s_{b}\stackrel{RD}{\rightarrow}s_{a}\}\\
\leq &\frac{1}{N_{S}}\sum_{N=1}^{N_{S}} \sum_{a=1}^{|{\mathcal{A}}_{y}|}P\{y_{a}\}\sum_{y_{b}\in{\mathcal{A}}_{a,N}}Q\left(\frac{d_{ab}^{(y)}}{\sqrt{2N_{0}}}\right)Q\left(\frac{d_{ba}^{(s)}}{\sqrt{2N_{0}}}\right)
\\= &\hat{P}''',
\end{split}
\end{equation}
where ${\mathcal{A}}_{a,N}=\{y_{b}|y_{b}\in {\mathcal{A}}_{y}, ({\mathbf{z}}_{a})_{N}\neq ({\mathbf{z}}_{b})_{N}\}$.
\subsubsection{case 2}
Transmission errors occur in $R\stackrel{\mathrm{TS}\: n}{\longrightarrow}D$. We assume that $x^{(n)}_{N_{r}}$, the $N$th element of ${\mathbf{x}}^{(n)}_{r}=[x^{(n)}_{1_{r}},...,x^{(n)}_{N_{S_{r}}}]^{T}$, is sentenced to $x'^{(n)}_{N_{r}}$ in $R\text{\bfseries{--}}D$ link at TS $n$. In order to recover $x^{(n)}_{N}$ simultaneously sent by sources at TS $n$, the symbol $x'^{(n+1)}_{N_{r}}$ detected by $D$ at TS $n+1$ should satisfy
\begin{equation}
\begin{split}
\label{36}
g(x'^{(n)}_{N_{r}},x'^{(n+1)}_{N_{r}})=g(x^{(n)}_{N_{r}},x^{(n+1)}_{N_{r}})=x^{(n)}_{N},
\end{split}
\end{equation}
where $x^{(n)}_{N_{r}}$ and $x^{(n+1)}_{N_{r}}$ are the correct symbols supposed to be detected by $D$ without errors. Hence, the probability of case 2 can be expressed as
\begin{equation}
\begin{split}
\label{37}
P_{case-2}=&P\{x^{(n)}_{N_{r}}\stackrel{RD}{\rightarrow}x'^{(n)}_{N_{r}}\}\cdot P\{x^{(n)}_{N}\stackrel{S-R-D}{\longrightarrow}x'^{(n+1)}_{N_{r}}\}\\
=&\frac{1}{N_{S}|{\mathcal{A}}_{s}|}\sum_{N=1}^{N_{S}} \sum_{a=1}^{|{\mathcal{A}}_{s}|}P\{s_{a}\stackrel{RD}{\rightarrow}s'_{a}\}\\
&\cdot\frac{1}{|{\mathcal{A}}_{s}|}
\sum_{b=1}^{|{\mathcal{A}}_{s}|}\sum_{k=1}^{|{\mathcal{A}}_{y}|}P\{y_{b}\stackrel{SR}{\rightarrow}y_{k}\}\sum_{s'_{k}} P\{s_{k}\stackrel{RD}{\rightarrow}s'_{k}\},
\end{split}
\end{equation}
where $s'_{a}$ is the correspondence to ${\mathbf{x}}'_{a_{r}}=[...,x'_{a_{N_{r}}},...]^{T}$. In $(\ref{37})$, $y_{b}=s_{a}+s_{b}$ and $s'_{k}$ is the correspondence to ${\mathbf{x}}'_{k_{r}}=[...,x'_{k_{N_{r}}},...]^{T}$, which satisfies $x_{N}=g(x'_{a_{N_{r}}},x'_{k_{N_{r}}})$. Replacing $P\{\cdots\}$ with $Q(\cdots)$, $P_{case-2}$ is given by
\begin{equation}
\begin{split}
\label{38}
P_{case-2}\leq&\frac{1}{N_{S}|{\mathcal{A}}_{s}|}\sum_{N=1}^{N_{S}} \sum_{a=1}^{|{\mathcal{A}}_{s}|}Q\left(\frac{d_{aa'}^{(s)}}{\sqrt{2N_{0}}}\right)\\
&\cdot\frac{1}{|{\mathcal{A}}_{s}|}
\sum_{b=1}^{|{\mathcal{A}}_{s}|}\sum_{k=1}^{|{\mathcal{A}}_{y}|}Q\left(\frac{d_{bk}^{(y)}}{\sqrt{2N_{0}}}\right)\sum_{s'_{k}} Q\left(\frac{d_{kk'}^{(s)}}{\sqrt{2N_{0}}}\right)\\
=&\hat{P}_{case-2}.
\end{split}
\end{equation}

According to $(\ref{25})$,$(\ref{27})$,$(\ref{35})$ and $(\ref{38})$, we define $\hat{P}_{e-RD}$, $\hat{P}_{e-SR}$, $\hat{P}'''$ and $\hat{P}_{case-2}$ as the bounds for $P_{e-RD}$, $P_{e-SR}$, $P'''$ and $P_{case-2}$ respectively by replacing $P\{\cdots\}$ with $Q(\cdots)$. Besides, $P_{case-1}=(1-P_{e-RD})\cdot P'''$. Let
\begin{equation}
\begin{split}
\label{41}
\hat{P}_{case-1}=(1-\hat{P}_{e-RD})\cdot \hat{P}'''.
\end{split}
\end{equation}
Based on the fact that $P''=P_{case-1}+P_{case-2}$, we obtain the lower bound for $T_{new}$:
\begin{equation}
\begin{split}
\label{42}
T_{new}\geq (1-\hat{P}_{e-SR})(1-\hat{P}_{e-RD})^{2}+\hat{P}_{case-1}+\hat{P}_{case-2},
\end{split}
\end{equation}
and the proof of "$\geq$" is similar to the derivation of CFNC-based scheme.

In AWGN channels without fading, the union bound for SEP, obtained from $(\ref{23})$, approaches the actual value. Therefore, the lower bounds can be the approximations of practical throughput, especially in high SNR regime. When SNR tends to infinity (i.e. $N_{0}$ tends to zero), the SEPs tend to zero and we can achieve
\begin{equation}
\label{43}
T_{CFNC}={\mathop {\lim }\limits_{N_{0} \to 0 }}\frac{1}{2}[(1-P'_{e-SR})(1-P_{e-RD})+P']=\frac{1}{2}
\end{equation}
\begin{equation}
\label{44}
\quad\:\:\: T_{new}= {\mathop {\lim }\limits_{N_{0} \to 0 }}(1-P_{e-SR})(1-P_{e-RD})^{2}+P''=1,
\end{equation}
from which the throughput of 1 sym/S/TS is verified.

\subsection{Practical system with Rayleigh fading channels}
Now we consider a more practical system where the channels are assumed to be flat Rayleigh fading. We have $h_{S_{i}R_{m}}\sim {\mathcal{CN}}(0,\sigma_{im}^{2}),h_{12},h_{21}\sim {\mathcal{CN}}(0,\sigma_{12}^{2}),h_{R_{m}D}\sim {\mathcal{CN}}(0,\sigma_{mD}^{2})$, $(i=1,2,...,N_{S},m=1,2)$.

We still firstly analyze the link $R\text{\bfseries{--}}D$, where the transmitted symbol is $h_{R_{m}D}\Theta^{T}{\mathbf{x}}_{r}$. Each vector's length in the constellation of $s$ is multiplied by $|h_{R_{m}D}|$ which
follows Rayleigh distribution. Therefore, $P\{s_{a}\rightarrow s_{b}|h_{R_{m}D}\}$ can be expressed as
\begin{equation}
\label{045}
P\{s_{a}\rightarrow s_{b}|h_{R_{m}D}\}\leq Q\left(\sqrt{\frac{|h_{R_{m}D}|^{2}d_{ab}^{2_{(s)}}}{2N_{0}}}\right).
\end{equation}
where $d_{ab}^{(s)}$ denotes the Euclidean distance between $s_{a}$ and $s_{b}$ in AWGN channel.
According to Chernoff bound [20], we acquire
\begin{equation}
\label{046}
Q\left(\sqrt{\frac{|h_{R_{m}D}|^{2}d_{ab}^{2_{(s)}}}{2N_{0}}}\right)\leq {\mathrm{exp}}\left(-\frac{|h_{R_{m}D}|^{2}d_{ab}^{2_{(s)}}}{4N_{0}}\right).
\end{equation}
Probability density function (pdf) of Rayleigh distribution is
\begin{equation}
\label{047}
p(v)=\left\{ \begin{array}{ll}
\large{\frac{v}{\sigma^{2}}{\mathrm{exp}}\left(-\frac{v^{2}}{2\sigma^{2}}\right)}&(0\leq v\leq \infty)\\
0&(v<0)
\end{array}\right .
\end{equation}
Besides, the phase of $h_{R_{m}D}$ follows uniform distribution on intervals $[0,2\pi]$ and is independent of the distribution of $|h_{R_{m}D}|$.
Hence, let $v_{mD}=|h_{R_{m}D}|$ and $\angle h_{R_{m}D}=\alpha$, we can acquire
\begin{equation}
\begin{split}
\label{048}
&P\{s_{a}\rightarrow s_{b}\}\leq {\mathrm{E}}_{h_{R_{m}D}}\left[{\mathrm{exp}}\left(-\frac{|h_{R_{m}D}|^{2}d_{ab}^{2_{(s)}}}{4N_{0}}\right)\right]\\
&= \int^{2\pi}_{0} \frac{1}{2\pi}\int^{\infty}_{0} \frac{v_{mD}}{\sigma^{2}_{mD}}{\mathrm{exp}}\left(-\frac{v_{mD}^{2}}{2\sigma^{2}_{mD}}-\frac{v_{mD}^{2}d_{ab}^{2_{(s)}}}{4N_{0}}\right)dv_{mD}d\alpha\\
&=\frac{2N_{0}}{2N_{0}+\sigma^{2}_{mD}d_{ab}^{2_{(s)}}}.
\end{split}
\end{equation}
Replacing $Q\left(\frac{d_{ab}^{(s)}}{\sqrt{2N_{0}}}\right)$ with $\frac{2N_{0}}{2N_{0}+\sigma^{2}_{mD}d_{ab}^{2_{(s)}}}$ in $(\ref{25})$, the SEP bound for $R_{m}\text{\bfseries{--}}D$ link is attainable. $P_{e-RD}=\frac{1}{2}(P_{e-R_{1}D}+P_{e-R_{2}D})$ owing to the equivalence of two relays.

Then, we consider the performance of $S\text{\bfseries{--}}R$ link. Regardless of noise, the signal received by relay is $\Theta^{T}{\mathbf{H}}_{SR}{\mathbf{x}}+\Theta^{T}h_{12}{\mathbf{x}}_{r}$. Let $\tilde{\mathbf{z}}={\mathbf{H}}_{SR}{\mathbf{x}}+h_{12}{\mathbf{x}}_{r}$ and $\tilde{y}=\Theta^{T}\tilde{\mathbf{z}}$. Then, $(\ref{27})$ can be rewritten as
\begin{equation}
\begin{split}
\label{049}
P_{e-SR}\leq\frac{1}{N_{S}}\sum_{k=1}^{N_{S}}k \sum_{a=1}^{|{\mathcal{A}}_{\tilde{y}}|}P\{\tilde{y}_{a}\}\left(\sum_{\begin{subarray}{c}b=1,\\||\tilde{{\mathbf{z}}}_{a}-\tilde{{\mathbf{z}}}_{b}||_{0}=k
\end{subarray}}^{|{\mathcal{A}}_{\tilde{y}}|}Q\left(\frac{d_{ab}^{(\tilde{y})}}{\sqrt{2N_{0}}}\right)\right),
\end{split}
\end{equation}
where $d_{ab}^{(\tilde{y})}$ denotes the Euclidean distance between $\tilde{y}_{a}$ and $\tilde{y}_{b}$. Since all channels are independent of each other, we have
\begin{equation}
\begin{split}
\label{050}
&Q\left(\frac{d_{ab}^{(\tilde{y})}}{\sqrt{2N_{0}}}\right)\leq \mathrm{E}_{{\mathbf{H}}_{SR},h_{12}}\left[\mathrm{exp}\left(-\frac{d_{ab}^{2_{(\tilde{y})}}}{4N_{0}}\right)_{{\mathbf{H}}_{SR},h_{12}}\right]\\
&=\int^{2\pi}_{0}\!\!\!\int^{\infty}_{0}\!\!\!\cdots\int^{2\pi}_{0}\!\!\!\int^{\infty}_{0}\!\!\frac{v_{1m}v_{2m}\cdots v_{N_{S}m}v_{12}}{\sigma_{1m}^{2}\sigma_{2m}^{2}\cdots\sigma_{N_{S}m}^{2}\sigma_{12}^{2}}\mathrm{exp}\Big(-\\&(\frac{v_{1m}^{2}}{2\sigma^{2}_{1m}}+
\cdots+\frac{v_{N_{S}m}^{2}}{2\sigma^{2}_{N_{S}m}}+\frac{v_{12}^{2}}{2\sigma^{2}_{12}}+\frac{d_{ab}^{2_{(\tilde{y})}}}{4N_{0}})\Big)dv_{12}d\alpha_{12}\\
&dv_{N_{S}m}d\alpha_{N_{S}m}\cdots dv_{1m}d\alpha_{1m},
\end{split}
\end{equation}
where ${\mathbf{H}}_{SR}=diag(h_{S_{1}R_{m}},h_{S_{2}R_{m}},...,h_{S_{N_{S}}R_{m}})$ and $|h_{S_{i}R_{m}}|=v_{im},|h_{12}|=v_{12}$ $(i=1,2,...,N_{S}\:\:m=1,2)$. The phases of channel coefficients are $\angle h_{S_{i}R_{m}}=\alpha_{im}$ and $\angle h_{12}=\alpha_{12}$. In addition,
\begin{equation}
\begin{split}
\label{051}
d_{ab}^{2_{(\tilde{y})}}=|\Theta^{T}({\mathbf{H}}_{SR}{\mathbf{x}}_{a}+h_{12}{\mathbf{x}}_{a_{r}}-{\mathbf{H}}_{SR}{\mathbf{x}}_{b}-h_{12}{\mathbf{x}}_{b_{r}})|^{2},
\end{split}
\end{equation}
where ${\mathbf{H}}_{SR}{\mathbf{x}}_{a}+h_{12}{\mathbf{x}}_{a_{r}}=\tilde{\mathbf{z}}_{a}$ and ${\mathbf{H}}_{SR}{\mathbf{x}}_{b}+h_{12}{\mathbf{x}}_{b_{r}}=\tilde{\mathbf{z}}_{b}$. Since all the $v_{im}$ and $\alpha_{im}$ are independent of each other, $(\ref{050})$ can be obtained through one by one integration. To reduce the complexity of integral, we make some simplifications, e.g., if the sources stay close during a period of time, we may claim the same channel coefficient for each of them. Besides, due to the large scale of constellation, we can only use $d_{a}^{min}$, rather than $d_{ab}^{(\tilde{y})}$, for each $\tilde{y}_{a}$ where $d_{a}^{min}$ is the minimum Euclidean distance between $\tilde{y}_{a}$ and other points in the constellation. Since $d_{a}^{min}$ mainly determines the SEP performance of $P_{e}\{\tilde{y}_{a}\}$, the mean error can be low enough. For any given ${\mathbf{x}}_{a},{\mathbf{x}}_{a_{r}},{\mathbf{x}}_{b}$ and ${\mathbf{x}}_{b_{r}}$, by replacing $Q\left(\frac{d_{ab}^{(\tilde{y})}}{\sqrt{2N_{0}}}\right)$ with $(\ref{050})$, the SEP bound for $S\text{\bfseries{--}}R$ link $(\ref{049})$ can be achieved by calculation.

The subsequent analysis of throughput is the same as that in AWGN channels, based on $P_{e-SR}$ and $P_{e-RD}$. The performance of our proposed scheme in Rayleigh fading channels through simulations is presented in the next section.

\section{Simulation}
\subsection{Simulation Setup}
We apply the transmission scheme to the uplinks in cellular network and the simulation parameters are configured according to the 3GPP LTE specifications [21][22] and specific parameters related to such transmission scheme as well, which are shown in Table V.
\begin{table}[ht]
\centering
\caption{Simulation parameters}
 \begin{tabular}{cc}
 \toprule
  Parameters & Value\\
  \midrule
  Modulation Mode & BPSK,QPSK \\
  Packet Size & 128 bits\\
  User Antenna Gain & 0 dBi\\
  $\Theta^{T}$ & LCF[14]\\
  Noise Power & -174 dBm/Hz\\
  Path Loss & $15.3+37.6{\mathrm{log}}_{10}d$\\
  Uplink Receiver Type  & MRC(Maximum Ratio Combining) \\
  \bottomrule
 \end{tabular}
\end{table}
\begin{figure}[h]
\centering
\includegraphics[width=0.48\textwidth]{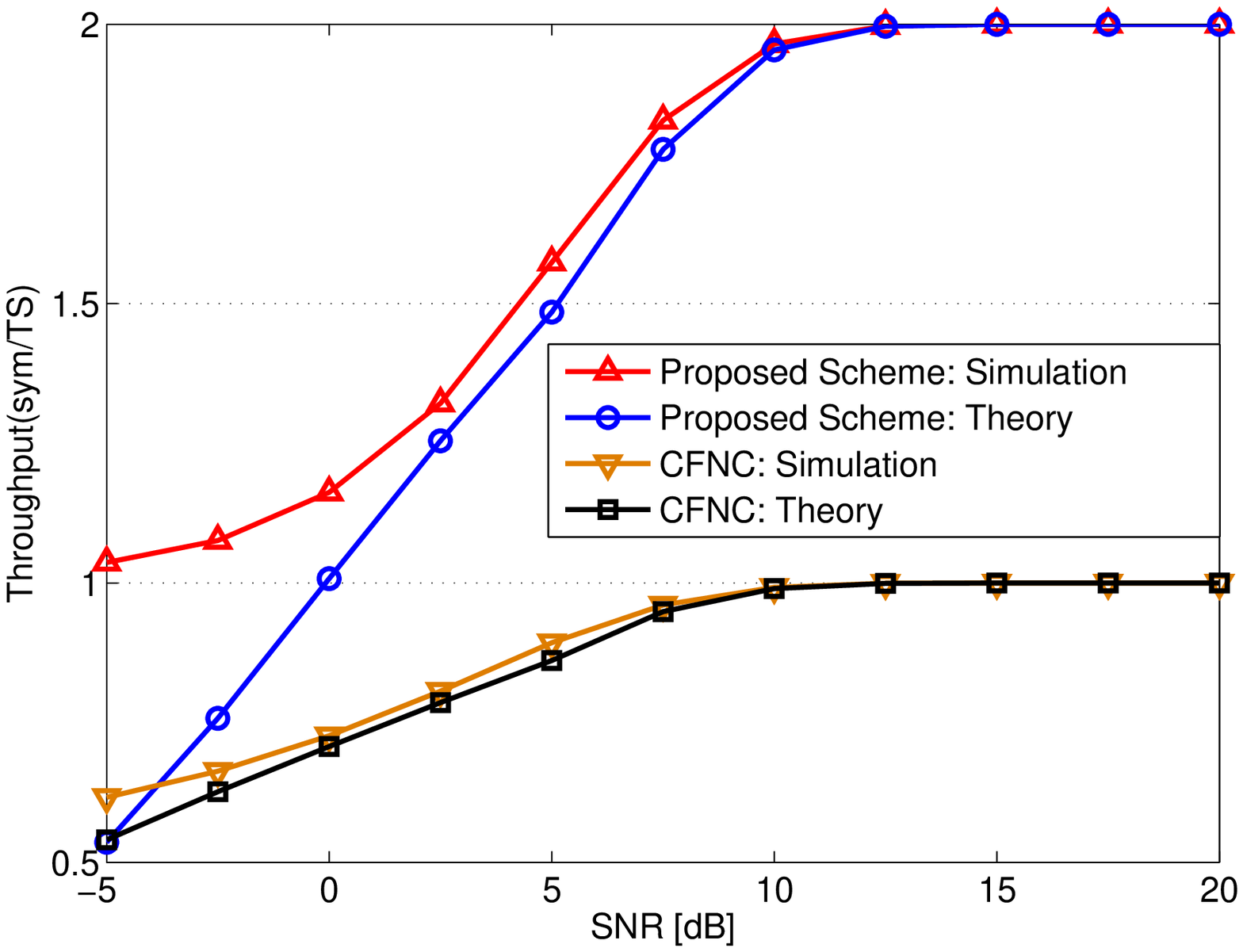}
\caption{Throughput for (2-2-1) system in AWGN channels}
\label{f5}
\end{figure}
\begin{figure}[h]
\centering
\includegraphics[width=0.48\textwidth]{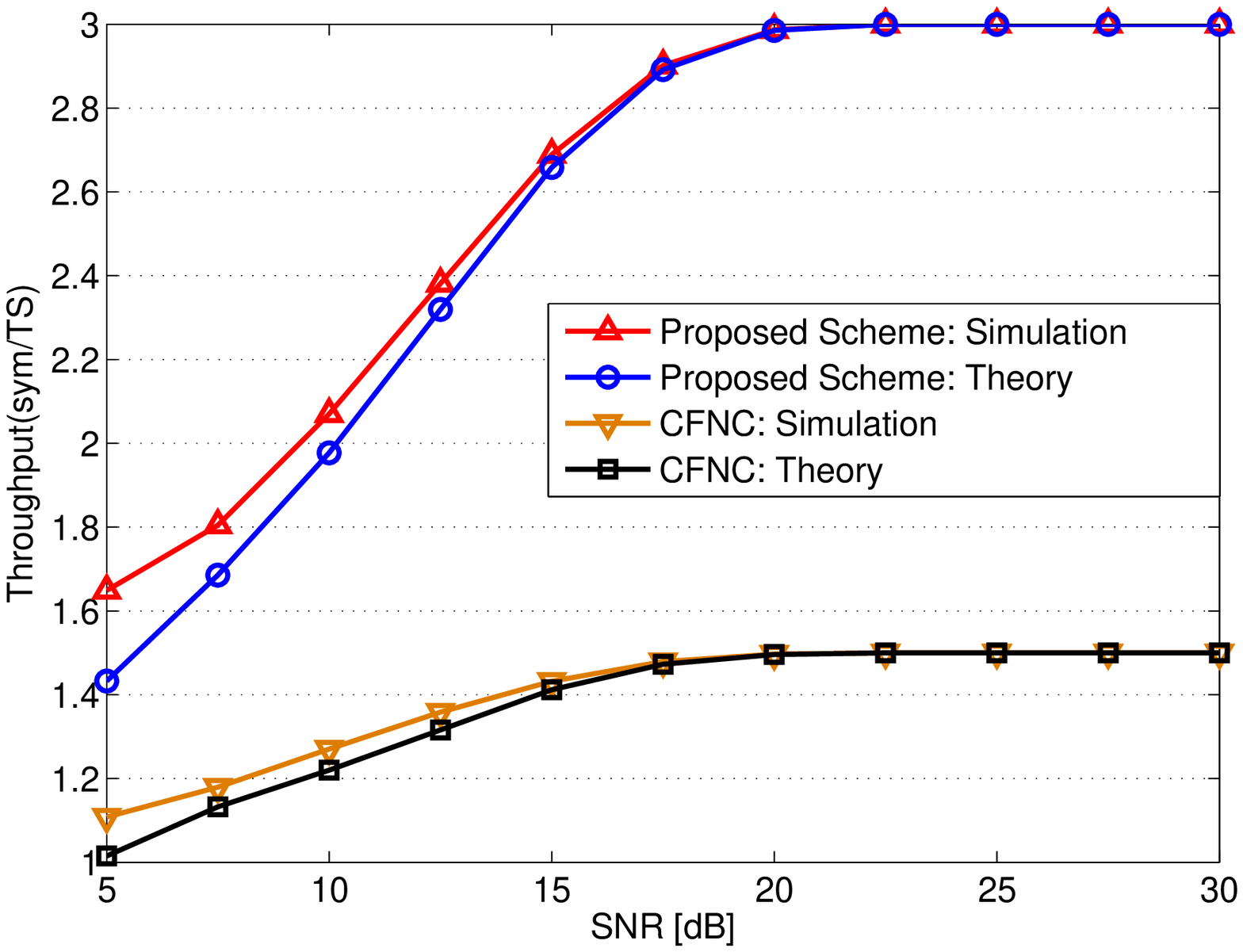}
\caption{Throughput for (3-2-1) system in AWGN channels}
\label{f5}
\end{figure}
\begin{figure}[h]
\centering
\includegraphics[width=0.48\textwidth]{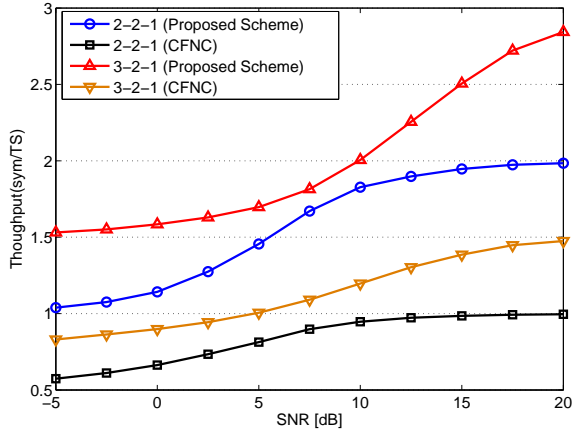}
\caption{Throughput performance in Rayleigh fading channels}
\label{f6}
\end{figure}
\begin{figure}[h]
\centering
\includegraphics[width=0.48\textwidth]{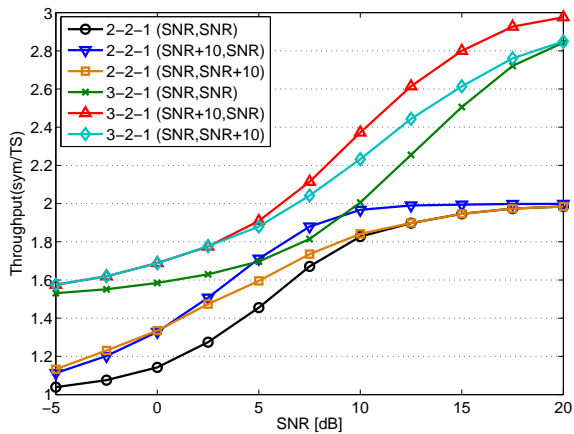}
\caption{Throughput comparison for various channel SNR settings}
\label{f7}
\end{figure}
\begin{figure}[h]
\centering
\includegraphics[width=0.48\textwidth]{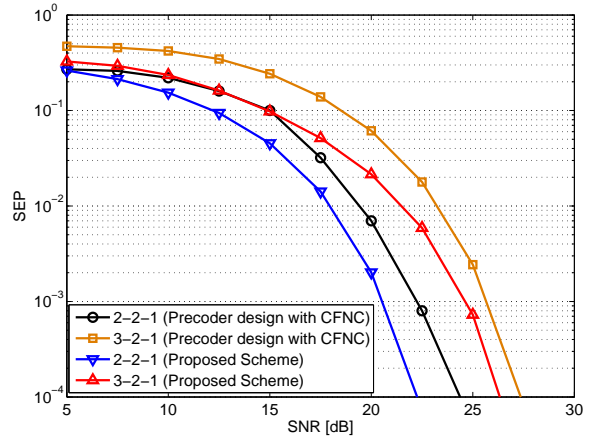}
\caption{SEP comparison in AWGN channels}
\label{f8}
\end{figure}
\begin{figure}[h]
\centering
\includegraphics[width=0.48\textwidth]{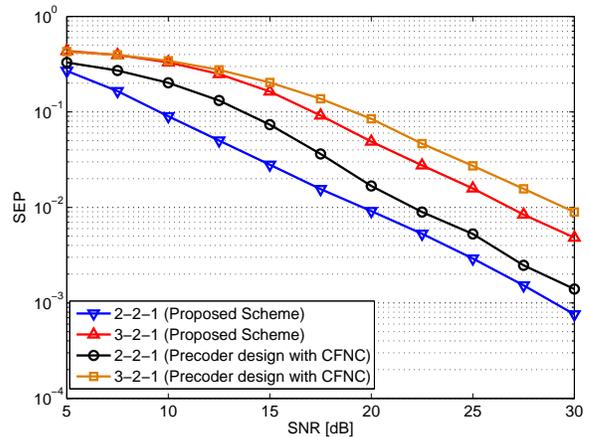}
\caption{SEP comparison in Rayleigh fading channels}
\label{f9}
\end{figure}
\begin{figure}[h]
\centering
\includegraphics[width=0.48\textwidth]{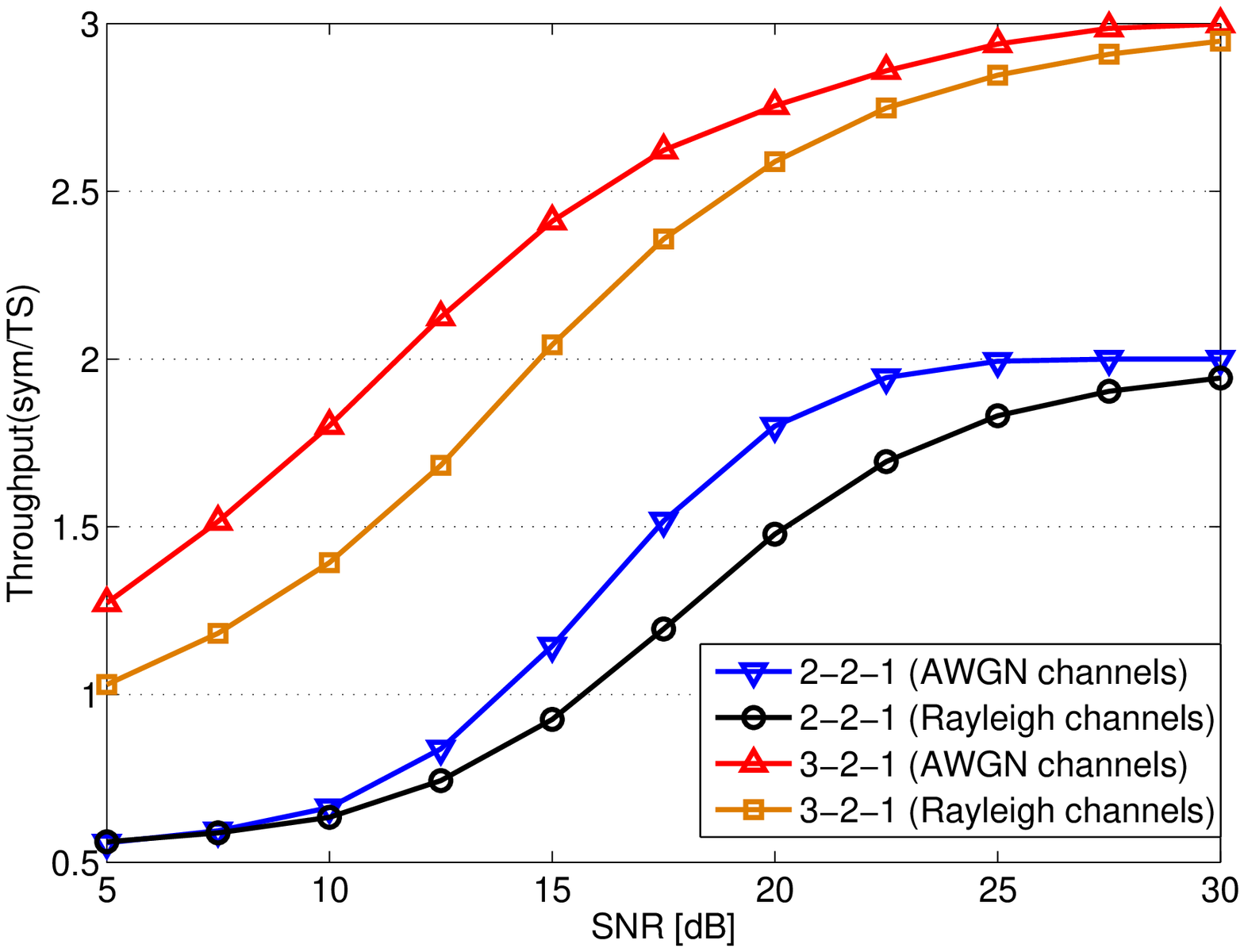}
\caption{Throughput performance adopting QPSK modulation}
\label{f10}
\end{figure}

Two scenarios of channel are considered: AWGN channel and Rayleigh fading channel. The design of $\Theta^{T}$ is the same as the description in section III. Through ML detection, symbols are detected at both relay nodes and destination. Assuming that the noise power spectral density $N_{0}$ remains unchanged during the entire transmission process, $Q\left(\frac{d_{ab}^{(s)}}{\sqrt{2N_{0}}}\right)$ is equivalent to $Q\left(\frac{d_{ab}'^{(s)}}{\sqrt{2}}\sqrt{\mathrm{SNR}}\right)$ where SNR is the received SNR and $d_{ab}'^{(s)}$ is the Euclidean distance between $s_{a}$ and $s_{b}$ in the constellation where $\mathrm{E}\{||s||^{2}=1\}$ (i.e. transmitted power is normalized to be 1). Given modulation mode, the number of the sources ($N_{S}$) and the design of $\Theta^{T}$, the throughput is only determined by SNR. We still assume $N_{0-SR}=N_{0-RD}=N_{0}$. Therefore, we mainly consider the performance of SEP and throughput in various scenarios for practical SNR values. Additionally, since the linear growth of $N_{S}$ leads to exponential growth of cardinality of $\Theta^{T}\mathbf{x}$ and $\Theta^{T}({\mathbf{x}}+{\mathbf{x}}_{r})$, which results in deterioration of SEP and throughput, we mainly consider the systems with 2-source and 3-source. To verify the improvement of the performance and illustrate the difference between 2-source and 3-source, the throughput in this section is defined as the number of successfully transmitted symbols from all sources per TS (sym/TS), rather than sym/S/TS.

\subsection{Results and Discussions}
Firstly, we compare the performance between proposed scheme and traditional CFNC scheme. CFNC [3] needs two TSs to complete one transmission. In our proposed scheme, with two relays alternatively forwarding the symbols to $D$, only $(L+1)$ TSs are required to transmit $L$ symbols. BPSK modulation is adopted for all nodes. We evaluate the throughput performance for ($N_{S}$,2,1) systems in AWGN channels.
In Fig.$\:$5 and Fig.$\:$6, the average throughput of proposed scheme based on two-path relay with IRIC in (2,2,1) and (3,2,1) systems is compared with that of traditional CFNC. Even when SNR is as low as -5 dB, the throughput of new scheme is higher than the upper limit of traditional CFNC. With the increase of SNR, the throughput of proposed scheme approaches the upper limit, which is 2 sym/TS in 2-source system while 3 sym/TS in 3-source network. Furthermore, we calculate the theoretical values of throughput based on $(\ref{30})$ and $(\ref{42})$, marked as Theory, and compare them with the simulation values under different SNR. To clearly investigate the tendency of the lower bound derived above, SNR varies from 5 dB to 30 dB due to large scale of constellation for 3-source system. It can be observed that the theoretical values and the simulation values are almost identical in high SNR regime, which verifies the derivation in Section V. Moreover, the theoretical values demonstrate that even the lower bounds achieve $N_{S}$ sym/TS, not to mention the actual values. Then, we consider a more practical system where the channels are assumed to be flat Rayleigh fading. Although the throughput decreases compared with that in AWGN channels due to channel fading, the new scheme can also reach the upper limit in high average SNR regions, as shown in Fig.$\:$7.

Secondly, throughput comparisons for various SNR settings in Rayleigh fading channels are given in Fig.$\:$8. The relay nodes always locate between sources and destination, so link $S\text{\bfseries{--}}R$ or $R\text{\bfseries{--}}D$ achieves better link quality than that of direct link $S\text{\bfseries{--}}D$. Link quality can be represented by SNR. Since $R$ is located either close to sources, close to $D$ or not close to both, we set corresponding SNRs $(\gamma_{SR},\gamma_{RD})$ in logarithmic-scale are (SNR$+10$ dB,SNR), (SNR,SNR$+10$ dB) and (SNR,SNR), respectively. According to the simulation results, the relays close to sources (SNR$+10$ dB,SNR) are supposed to have high-priority in relay selection, especially in high SNR regime. Furthermore, better link quality of $S\text{\bfseries{--}}R$ requests lower transmitted power at sources, which reduces inter-cell interference caused by cell-edge users.

A novel transmission scheme based on precoder design is proposed in [7], which also achieves symbol rate as high as 1 sym/S/TS regardless of noise and interference. During traditional transmission process, the symbol is transmitted one by one separately. In [7], source's symbol transmitted in channel is the superposition of two original modulated symbols, which doubles the ideal throughput. Therefore, we evaluate the performance of symbol error probability (SEP) between our new scheme and the scheme proposed in [7], shown in Fig.$\:$9 and Fig.$\:$10. Although the scheme based on precoder design is free from IRI, half of the symbols are detected based on $S\text{\bfseries{--}}D$ link in [7]. However, the link quality of $S\text{\bfseries{--}}D$ is worse than that of $S\text{\bfseries{--}}R\text{\bfseries{--}}D$, which apparently deteriorates the whole performance. Therefore, in either AWGN or Rayleigh fading channels, it can be observed that our proposed scheme achieves better SEP performance than what is proposed in [7].

All the simulations above are based on BPSK modulation. To check modulation other than BPSK, we test the throughput performance of proposed scheme adopting QPSK modulation. From Fig.$\:$11, although many transmission errors occur under low SNR due to large scale of the constellation, the throughput can also achieve the upper limit when SNR is large enough in either AWGN or Rayleigh fading channels.


In general, the simulation results have verified that throughput of the proposed scheme in this paper is able to reach the upper limit (1 sym/S/TS). Compared with other method which can also attain 1 sym/S/TS in $N_{S}-$source system, our proposed scheme achieves better performance.

\section{Conclusion}
In this paper, we propose a novel cooperative communication scheme based on two-path relay with IRIC to improve throughput in multi-source wireless networks. The two-path successive relay scheme ensures that the source continuously sends symbol to the two relay nodes alternatively at any TS. Deploying CFNC, the single-source system with two-path relay is generalized to multi-source system, from which all the sources are able to broadcast their own symbols in the same TS on the same RBs. In order to deal with the IRI caused by relays, PNC is applied at relay nodes and IRI is successfully canceled at destination. Theoretical analysis is presented to verify the performance of proposed scheme. Lower bounds for throughput in AWGN channels and Rayleigh fading channels are given to estimate throughput of such system. The simulation results in various scenarios illustrate that this new scheme achieves higher throughput than that of other schemes in either AWGN or Rayleigh fading channels.

\section*{Appendix}
{\em The proof for $(\ref{30})$}:

Through $(\ref{29})$, we define $\hat{P}'$ to be

\begin{equation}
\begin{split}
\label{45}
\hat{P}'=\frac{1}{N_{S}|{\mathcal{A}}_{s}|}\sum_{N=1}^{N_{S}} \sum_{a=1}^{|{\mathcal{A}}_{s}|}\sum_{s_{b}\in{\mathcal{A}}_{a,N}}Q\left(\frac{d_{ab}^{(s)}}{\sqrt{2N_{0}}}\right)Q\left(\frac{d_{ba}^{(s)}}{\sqrt{2N_{0}}}\right).
\end{split}
\end{equation}
Therefore, we have $1-P_{e-RD}\geq 1-\hat{P}_{e-RD}$ and $P'\leq \hat{P}'$. Due to the fact that $a\geq b$ and $c\leq d$ cannot directly lead to $a+c\geq b+d$, our purpose is to verify that $T_{CFNC}\geq \frac{1}{2}[(1-\hat{P}_{e-RD})^{2}+\hat{P}']$.

The throughput $T_{CFNC}$ can be rewritten as
\begin{equation}
\begin{split}
\label{46}
T_{CFNC}=&\frac{1}{2}[1-2P_{e-RD}+P_{e-RD}^{2}+P']\\
=&\frac{1}{2}[1-P_{e-RD}(1-P_{e-RD})+(P'-P_{e-RD})],
\end{split}
\end{equation}
where
\begin{equation}
\begin{split}
\label{47}
-P_{e-RD}(1-P_{e-RD})\geq -\hat{P}_{e-RD}(1-\hat{P}_{e-RD}).
\end{split}
\end{equation}
Besides, let $P_{e-RD}+\Delta_{RD}=\hat{P}_{e-RD}$, $P_{e-RD}^{2}+\Delta'_{RD}=\hat{P}_{e-RD}^{2}$ and $P'+\Delta_{P'}=\hat{P'}$ $(0\leq \Delta_{RD}, \Delta'_{RD}, \Delta_{P'}\leq 1)$, we have
\begin{equation}
\begin{split}
\label{48}
-P_{e-RD}(1-P_{e-RD})=& -\hat{P}_{e-RD}(1-\hat{P}_{e-RD})\\&+\Delta_{RD}-\Delta'_{RD}.
\end{split}
\end{equation}
From $(\ref{47})$ and $(\ref{48})$, we have $\Delta_{RD}-\Delta'_{RD}\geq 0$. Moreover, note that $P_{e-RD}^{2}$ consists of all the cases when transmission errors occur, i.e. $x_{a}\stackrel{SR}{\rightarrow}x_{b}\stackrel{RD}{\rightarrow}x_{k}\:(x_{a}\neq x_{b},x_{b}\neq x_{k} )$, which contains $P':\:x_{a}\stackrel{SR}{\rightarrow}x_{b}\stackrel{RD}{\rightarrow}x_{a}\:(x_{a}\neq x_{b})$. Therefore, it is obvious that $\Delta'_{RD}\geq \Delta_{P'}$, so is $\Delta_{RD}-\Delta_{P'}\geq 0$. Hence, the second term of $(\ref{46})$:
\begin{equation}
\label{49}
P'-P_{e-RD}=\hat{P'}-\hat{P}_{e-RD}+\Delta_{RD}-\Delta_{P'}\geq \hat{P'}-\hat{P}_{e-RD},
\end{equation}
together with $(\ref{47})$, the lower bound for $T_{CFNC}$ is obtained by
\begin{equation}
\begin{split}
\label{50}
T_{CFNC}=\frac{1}{2}[(1-P_{e-RD})^{2}+P']\geq \frac{1}{2}[(1-\hat{P}_{e-RD})^{2}+\hat{P}'].
\end{split}
\end{equation}
The lower bound for $T_{CFNC}$ in $(\ref{30})$ is verified. Furthermore, the lower bound for $T_{new}$ can also be verified through similar method presented above.

\section{Acknowledgement}
The authors would like to thank the editor and anonymous reviewers for their valuable suggestions that significantly improve the quality of this paper. Thanks also to Prof. Wenyi Zhang(USTC), Prof. Chi Zhang(USTC), Dr. Hao Yue(University of Florida), Ph.D. Candidate Hao Tang(USTC) and Ph.D. Candidate Jinlin Peng(USTC) for helpful discussions and valuable suggestions. This work was supported by the National Natural Science Foundation of China (No. 60903216 and No. 61170231) and the National S\&T Major Project of China(No. 2010ZX03003-002 and No. 2011ZX03005-006).

\epsfysize=3.2cm
\addtolength{\itemsep}{-1em}
\setlength{\itemsep}{-0.5mm}
\begin{IEEEbiography}[{\includegraphics[width=1.2in,height=1.25in,clip,keepaspectratio]{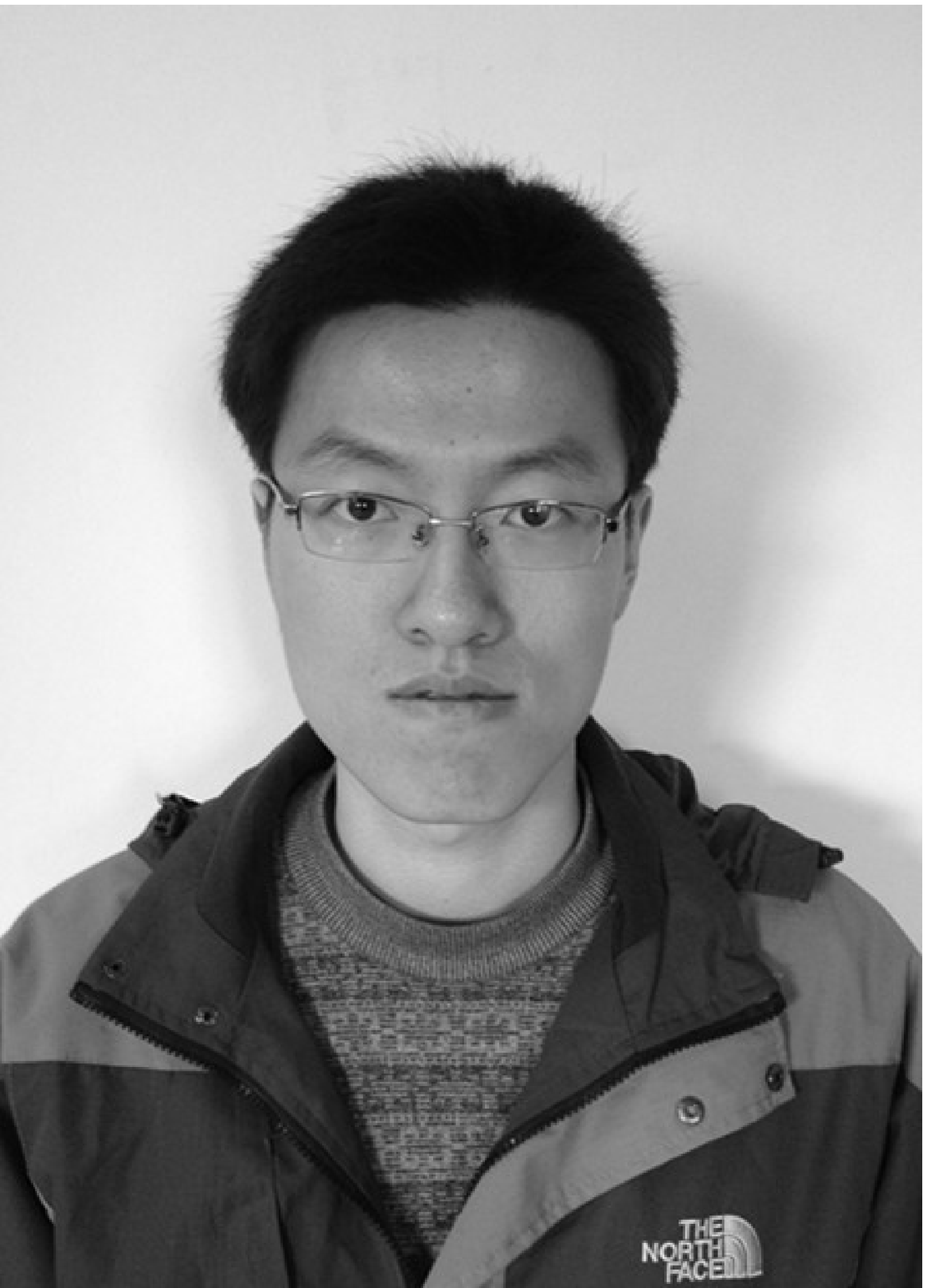}}]{Hao Lu}
was born in 1989. He received his B.S. degree at the Department of
Electronic Engineering and Information Science
(EEIS) from
University of Science and Technology of
China (USTC), Hefei, China in 2011. Now,
he is pursuing his Ph.D. degree of Communication
and Information Systems at EEIS from the same university. His research interests
lie in relay-based cooperative communications and network coding.
\end{IEEEbiography}


\begin{IEEEbiography}[{\includegraphics[width=1in,height=1.25in,clip,keepaspectratio]{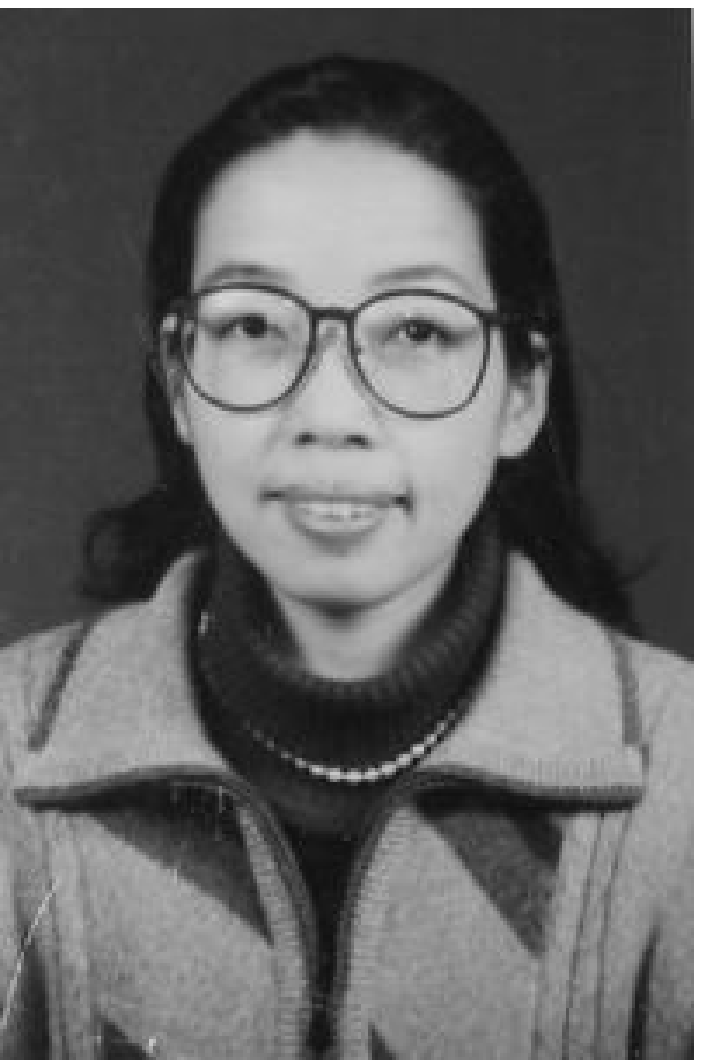}}]{Peilin Hong}
was born in 1961. She received her B.S. and M.S. degrees from the
Department of Electronic Engineering and Information Science (EEIS),
University of Science and Technology of China (USTC), in 1983 and
1986. Currently, she is a Professor and Advisor for Ph.D. candidates
in the Department of EEIS, USTC. Her research interests include
next-generation Internet, policy control, IP QoS, and information
security. She has published 2 books and over 100 academic papers in
several journals and conference proceedings.
\end{IEEEbiography}

\begin{IEEEbiography}[{\includegraphics[width=1in,height=1.25in,clip,keepaspectratio]{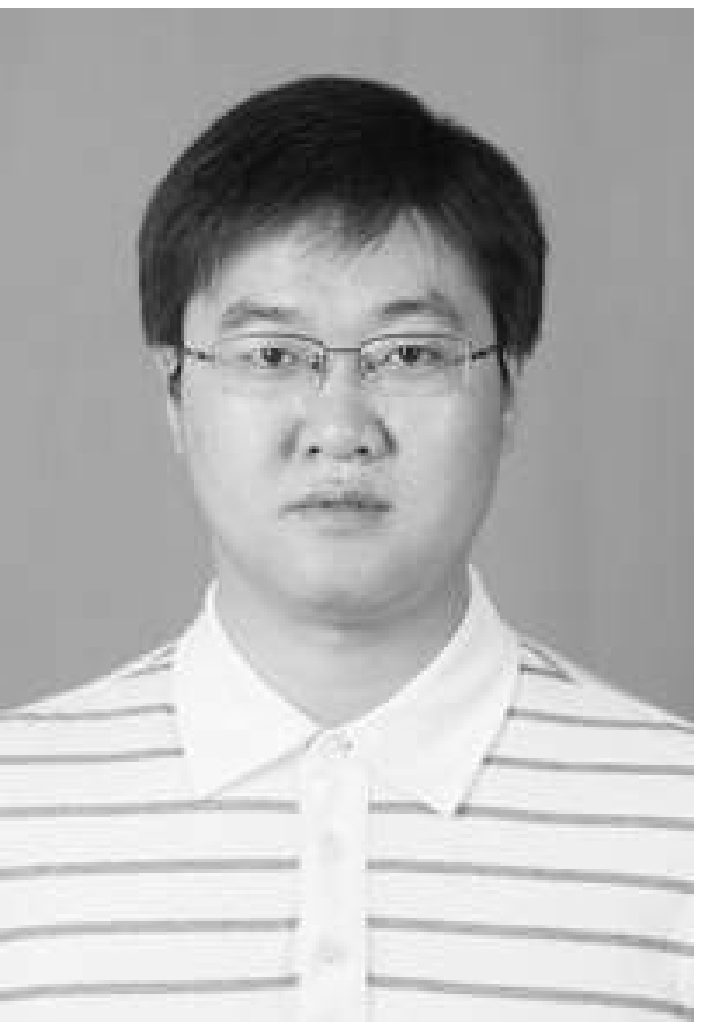}}]{Kaiping Xue}
was born in 1980. He graduated with a B.S. degree from the
Department of Information Security, University of Science and Technology of China
(USTC), in 2003 and received a Ph.D. degree from the Department of Electronic Engineering and Information
Science (EEIS), USTC, in 2007. Currently, he is an Associate Professor in the Department of
Information Security and Department of EEIS, USTC. His research
interests include next-generation Internet, distributed networks and
network security.
\end{IEEEbiography}


\begin{thebibliography}{1}





\bibitem{IEEEhowto:kopka}
S.~Zhang, S.~C.~Liew and P.~P.~Lam, "Hot topic: Physical-layer network noding," in {\em Proc. Intl. Conf. Mobile Computing \& Netw.}, Sep.~23-26, 2006, pp.~358-365.

\bibitem{IEEEhowto:kopka}
Y.~Chen, S.~Kishore and J.~Li, "Wireless diversity through network coding," in {\em Proc. Wireless Commun. \& Netw. Conf.}, Apr.~3-6, 2006, pp.~1681-1686.

\bibitem{IEEEhowto:kopka}
T.~Wang and G.~B.~Giannakis, "Complex field network coding for multiuser cooperative communications," {\em IEEE J. Sel. Areas Commun.}, vol. 26, no. 3, pp.~561-571, Apr. 2008.

\bibitem{IEEEhowto:kopka}
J.~Li, W.~Chen and X.~Wang, "Complex field network coding for wireless cooperative multicast flows," in {\em Proc. Global Telecommun. Conf.}, Nov.30-Dec.4, 2008, pp.~4604-4608.


\bibitem{IEEEhowto:kopka}
T.~Wang and G.~B.~Giannakis, "Capacity scaling of wireless networks with complex field network coding," {\em J. Commun.}, vol 4, no. 11, pp.~830-840, Dec. 2009.


\bibitem{IEEEhowto:kopka}
G.~Li, A.~Cano, J.~Gomez-Vilardebo, G.~B.~Giannakis and A.~I.~Perez-Neira, "High-throughput multi-source cooperation Via complex-field network coding," {\em IEEE Trans. Wireless Commun.}, vol 10, no. 5, pp.~1606-1617, May 2011.


\bibitem{IEEEhowto:kopka}
Y.~Ding, H.~M.~Kwon and K.~Lee, "Precoder design for amplify and forward relaying with complex field network coding," {\em IEEE Trans. Signal Process.}, vol 59, no. 6, pp.~2988-2994, June 2011.

\bibitem{IEEEhowto:kopka}
T.~Oechtering and A.~Sezgin, "A new cooperative transmission scheme using the space-time delay code," in {\em Proc. ITG Workshop Smart Antenna}, Mar.~18-19, 2004, pp.~41-48.

\bibitem{IEEEhowto:kopka}
A.~Ribeiro, X.~Cai and G.~B.~Giannakis, "Opportunistic multipath for bandwidth-efficient cooperative networking," in {\em Proc. Intl. Conf. Acoustics Speech and Signal Process.}, May 2004, pp.~549-552.


\bibitem{IEEEhowto:kopka}
C.~Luo, Y.~Gong and F.-C.~Zheng, "Interference cancellation in two-path successive relay system with network coding," in {\em Proc. Personal Indoor \& Mobile Radio Commun.}, Sept.~26-30, 2010, pp.~465-469.


\bibitem{IEEEhowto:kopka}
C.~Luo, Y.~Gong and F.-C.~Zheng, "Full interference cancellation for two-path relay cooperative networks," {\em IEEE Trans. Veh. Technol.}, vol. 60, no. 1, pp.~343-347, Jan. 2011.

\bibitem{IEEEhowto:kopka}
L.~Sun, T.~Zhang and H.~Niu, "Inter-relay interference in two-path digital relaying systems: detrimental or beneficial?," {\em IEEE Trans. Wireless Commun.}, vol. 10, no. 8, pp.~2468-2473, Aug. 2011.


\bibitem{IEEEhowto:kopka}
C.-C.~Chu, H.-C.~Wang and C.-L.~Wang, "Suboptimal Power Allocation for a Two-Path Successive Relay System with Full Interference Cancellation," in {\em Proc. Veh. Technol. Conf.-Spring}, May.~6-9 2012, pp.~1-5.



\bibitem{IEEEhowto:kopka}
Y.~Gong, C.~Luo and Z.~Chen, "Two-Path Succussive Relaying With Hybrid Demodulate and Forward," {\em IEEE Trans. Veh. Technol.}, vol.61, no.5, pp.~2044-2053, June 2012.



\bibitem{IEEEhowto:kopka}
IEEE Std. 802.11n-2009, "Part11: Wireless LAN Medium Access Control (MAC) and Physical Layer (PHY) Specifications: Enhancements for Higher Throughput," Oct. 2009.


\bibitem{IEEEhowto:kopka}
Y.~Xin, Z.~Wang and G.~B.~Giannakis, "Space-time diversity systems based on linear constellation precoding," \emph{IEEE Trans. Wireless Commun.}, vol. 2, no. 2, pp.~294-309, Mar. 2003.

\bibitem{IEEEhowto:kopka}
J.~Wang and X.~Liu, "Improvement of Complex Field Network Coding for Cooperative Networks," in \emph{Proc. Intl. Conf. Commun. \& Intelligence Inf. Secur.}, Oct.~13-14, 2010, pp.~176-179.

\bibitem{IEEEhowto:kopka}
3GPP standardization, "Physical Channels and Modulation," {\em TS 36.211, v11.0.0}, Sept. 2012.

\bibitem{IEEEhowto:kopka}
R.~E.~Ziemer and R.~L.~Peterson, {\em Introduction to Digital Communication}. Maxwell Macmillan International, 1992.

\bibitem{IEEEhowto:kopka}
J~G~Proakis. {\em Digital Communications}, $3^{rd}$ ed. McGraw-Hill, 1993.

\bibitem{IEEEhowto:kopka}
R4-092042, "Simulation assumptions and parameters for FDD HeNB RF requirements," {\em 3GPP TSG RAN WG4 (Radio) Meeting \#51}, May~4-8, 2009.

\bibitem{IEEEhowto:kopka}
3GPP standardization, "Radio Frequency(RF) system scenarios," {\em TR 36.942, v9.0.1}, Dec. 2010.
\end{thebibliography}
\end{document}